\documentclass[prl, twocolumn, superscriptaddress, nofootinbib, notitlepage, floatfix]{revtex4-2}
\usepackage{xcolor}
\usepackage{etoolbox}
\usepackage{graphicx}
\usepackage{wrapfig}
\usepackage{amsmath}
\usepackage{amsfonts}
\usepackage{amssymb}
\usepackage{multirow}
\usepackage{slashed}
\usepackage{physics}
\usepackage[colorlinks=true, linkcolor=blue, citecolor=blue, urlcolor=blue]{hyperref}

\usepackage{array, hhline}
\usepackage{ulem}

\definecolor{darkgreen}{rgb}{0,0.5,0}

\DeclareRobustCommand{\fig}[1]{Fig.~\ref{fig:#1}}

\DeclareRobustCommand{\tb}[1]{Table~\ref{tb:#1}}

\DeclareRobustCommand{\refcite}[1]{Ref.~\cite{#1}}
\DeclareRobustCommand{\refcites}[1]{Refs.~\cite{#1}}
\newcommand\bets{\begin{table*}}
\newcommand\eets[1]{\label{tb:#1}\end{table*}}

\newcommand{\rowheight}[1]{\renewcommand{\arraystretch}{#1}}
\newcolumntype{C}[1]{>{\centering\arraybackslash}m{#1}}

\begin{document}

%%%%%%%%%%%%%%%%%%%%%%%%%%%%%%%%%%%%%%%%%%%%%%%%%%%%%%%%%%%%%%%%%%%%%%
\title{QCD Predictions for Meson Electromagnetic Form Factors at High Momenta:\\ Testing Factorization in Exclusive Processes}
%%%%%%%%%%%%%%%%%%%%%%%%%%%%%%%%%%%%%%%%%%%%%%%%%%%%%%%%%%%%%%%%%%%%%%

%%%%%%%%%%%%%%%%%%%%%%%%%%%%%%%%%%%%%%%%%%%%%%%%%%%%%%%%%%%%%%%%%%%%%%

\author{Heng-Tong Ding}
\affiliation{Key Laboratory of Quark \& Lepton Physics (MOE) and Institute of Particle Physics, Central China Normal University, Wuhan 430079, China}

\author{Xiang Gao}
\email{gaox@anl.gov}
\affiliation{Physics Division, Argonne National Laboratory, Lemont, IL 60439, USA}

\author{Andrew D. Hanlon}
\affiliation{Physics Department, Brookhaven National Laboratory, Upton, New York 11973, USA}

\author{Swagato Mukherjee}
\affiliation{Physics Department, Brookhaven National Laboratory, Upton, New York 11973, USA}

\author{Peter Petreczky}
\affiliation{Physics Department, Brookhaven National Laboratory, Upton, New York 11973, USA}

\author{Qi Shi}
\email{qshi1@bnl.gov}
\affiliation{Key Laboratory of Quark \& Lepton Physics (MOE) and Institute of Particle Physics, Central China Normal University, Wuhan 430079, China}
\affiliation{Physics Department, Brookhaven National Laboratory, Upton, New York 11973, USA}

\author{Sergey Syritsyn}
\affiliation{RIKEN-BNL Research Center, Brookhaven National Laboratory, Upton, New York 11973}
\affiliation{Department of Physics and Astronomy, Stony Brook University, Stony Brook, New York 11790}

\author{Rui Zhang}
\affiliation{Physics Division, Argonne National Laboratory, Lemont, IL 60439, USA}

\author{Yong Zhao}
\affiliation{Physics Division, Argonne National Laboratory, Lemont, IL 60439, USA}

\begin{abstract}

We report the first lattice QCD computation of pion and kaon electromagnetic form factors, $F_M(Q^2)$, at large momentum transfer up to 10 and 28 $\mathrm{GeV}^2$, respectively. Utilizing physical masses and two fine lattices, we achieve good agreement with JLab experimental results at $Q^2 \lesssim 4~\mathrm{GeV}^2$. For $Q^2 \gtrsim 4~\mathrm{GeV}^2$, our results provide \textit{ab-initio} QCD benchmarks for the forthcoming experiments at JLab 12 GeV and future electron-ion colliders. We also test the QCD collinear factorization framework utilizing our high-{$Q^2$} form factors at 
next-to-next-to-leading order in perturbation theory, which relates the form factors to the leading Fock-state meson distribution amplitudes. Comparisons with independent lattice QCD calculations using the same framework demonstrate, within estimated uncertainties, the universality of these nonperturbative quantities.

\end{abstract}

\date{\today}     
\maketitle

%%%%%%%%%%%%%%%%%%%%%%%%%%%%%%%%%%%%%%%%%%%%%%%%%%%%%%%%%%%%%%%%%%%%%%
\textit{Introduction.} 
Elastic electron hadron scattering can be described in terms of electromagnetic form factors (EMFF), which 
characterize the charge distribution inside hadrons~\cite{Epelbaum:2022fjc}. The EMFF 
of the nucleon has been studied experimentally for many decades and provided the first 
glimpse of the complex internal structure of the nucleon~\cite{RevModPhys.30.482}, while the EMFF of pion and kaon, the Goldstone bosons of QCD, are much less known. On the other hand, the
study of the pion and kaon EMFF is important for at least two reasons. It has been argued that 
the pion and kaon form factors are important for understanding the dynamical generation of hadron
masses in QCD~\cite{Cui:2020dlm, Roberts:2021nhw}, and they are closely related
to the light-front wave functions of the pseudo-scalar mesons, see, e.g.,
\refcite{Alberg:2024svo}.
Studies of the pion and kaon electromagnetic form
factors at large momentum transfer, $Q^2$, and, more generally, of the hard exclusive processes are needed for
a more complete understanding of the partonic structure of the hadrons.
The partonic picture of hadrons is well established through the study of inclusive processes, but we need to see that this picture
also works in exclusive processes for an unambiguous interpretation of partons as the right degrees of freedom at short distances, see discussions in Refs. ~\cite{Brambilla:2014jmp,Huber:2008id}.
The QCD factorization for hard exclusive processes was proposed a long time ago~\cite{Farrar:1979aw, Lepage:1980fj}, but it is difficult to be tested experimentally due to the lack of experimental data. One clean exclusive process provided in recent years is the pion transition form factors, where large momentum transfer up to $40~\rm GeV^2$ is available~\cite{BaBar:2009rrj, Belle:2012wwz}, allowing its factorization into the distribution amplitudes~\cite{Braun:2021grd, Gao:2021iqq}. The EMFF, on the other hand, offer another valuable opportunity of testing the QCD factorization and is crucial for examining the universality of the factorization.

Direct experimental measurements of the pion EMFF are available only for very small $Q^2$, well below
$1~{\rm GeV}^2$~\cite{Dally:1981ur, Dally:1982zk, Amendolia:1984nz, Amendolia:1986wj}. It is possible to determine the 
pion form factor at higher $Q^2$ through pion electro-production off nucleons, but such determination
comes with some
model dependence~\cite{Huber:2008id}. However, also the measurements using this method do not extend to high enough
values of $Q^2$.
Measuring the kaon EMFF is even more challenging~\cite{Brauel:1979zk, Dally:1980dj, Amendolia:1986ui, Carmignotto:2018uqj}. Studies of the EMFF of the pion and kaon with high $Q^2$ up to $\sim$ 6 $\rm GeV^2$ are underway at the ongoing JLAB 12 GeV program~\cite{Dudek:2012vr, Arrington:2021alx}, and their measurements in an extended range of $Q^2\sim$ 9$-$40 $\rm GeV^2$ are planned at the future Electron-Ion Collider (EIC) facility~\cite{Arrington:2021biu} and Electron-ion collider in China (EicC)~\cite{Anderle:2021wcy}.

At present, lattice QCD is the only nonperturbative method that can directly predict EMFF without any model dependence, 
and the results can also be systematically improved. 
Therefore, first-principle calculations on a lattice can provide benchmark QCD 
predictions for comparison with experiments. Existing lattice calculations of the pion~\cite{QCDSFUKQCD:2006gmg, Boyle:2008yd, JLQCD:2009ofg, Bali:2013gya, Fukaya:2014jka, Colangelo:2018mtw, Wang:2020nbf, Gao:2021xsm, Koponen:2015tkr,ETM:2017wqc} and kaon~\cite{Kaneko:2010ru, Alexandrou:2021ztx} EMFF
are restricted to the low $Q^2$ region. Calculations of pion EMFF with $Q^2$ up to 6 $\rm GeV^2$ were performed in \refcite{QCDSF:2017ssq} using the Feynman-Hellmann method, albeit with large uncertainties. There  also exist lattice QCD calculations of the EMFF at large momentum transfer for pseudo-scalar mesons with strange-antistrange quark and charm-anticharm quarks \cite{Koponen:2017fvm,Davies:2018zav}. In this work, we study the pion and kaon EMFF with large momentum transfers $Q^2$ up to 10 $\rm GeV^2$ and 28 $\rm GeV^2$, respectively, using optimized boosted sources for large momenta in both the initial and final states. The calculations are performed directly at the physical point. Moreover, with independent lattice QCD calculations of the pion and kaon light-cone distribution amplitudes~\cite{RQCD:2019osh,Detmold:2021qln,LatticeParton:2022zqc, Gao:2022vyh, Baker:2024zcd, Cloet:2024vbv}, as well as the state-of-the-art perturbative input at next-to-next-to-leading order (NNLO)~\cite{Chen:2023byr}, we are able to verify the collinear leading-twist QCD factorization of the EMFF at such high $Q^2$~\cite{Farrar:1979aw, Lepage:1980fj} for the first time, thus demonstrating the universality within these nonperturbative quantities.

%%%%%%%%%%%%%%%%%%%%%%%%%%%%%%%%%%%%%%%%%%%%%%%%%%%%%%%%%%%%%%%%%%%%%%
\textit{Lattice QCD calculations of the form factors.} 
We extract the bare meson form factors from meson two-point and meson-current three-point correlation functions. 
We use two lattice QCD gauge ensembles generated by the HotQCD collaboration~\cite{Bazavov:2019www} 
with 2+1 flavors of highly improved staggered quarks (HISQ)~\cite{Follana:2006rc}.
These ensembles are defined on $L_s^3 \times L_t=64^3\times64$ lattices with spacings $a = 0.076$ fm and 0.04 fm. 
The strange quark mass $m_s$ for these ensembles is set to the physical value while the light quark masses are set to $m_s/27$ and $m_s/20$, respectively.
The light quark masses correspond to the pion masses of 
140 MeV and 160 MeV for the coarser and the finer lattice, respectively.
For the valence quarks, we use the Wilson-clover action with 1-step hypercubic (HYP) smeared~\cite{Hasenfratz:2001hp} 
gauge links and with tree-level tadpole improved coefficients 
$c_{sw}$ = 1.0372 and 1.02868~\cite{Gao:2022iex, Gao:2020ito} for the coarser and the finer lattices, 
respectively, which were determined from the smeared plaquette averages. 
The valence quark masses are tuned so that the pion/kaon masses 
are 140(1)/498(1) MeV for the $a$ = 0.076 fm ensemble and 134(3)/497(4) MeV for the $a$ = 0.04 fm ensemble. 
We use the QUDA multigrid algorithm~\cite{Brannick:2007ue, Clark:2009wm, Babich:2011np, Clark:2016rdz} for the Wilson-Dirac operator inversions to calculate the quark propagators. All Mode Averaging (AMA) technique~\cite{Shintani:2014vja} is employed to increase the statistics.

In order to obtain the bare matrix elements of the ground state, 
we need to compute the two-point functions to extract the energy spectra and get the overlap amplitudes,
\begin{align}
C_{\rm 2pt}(\mathbf{P},t_s)=\left\langle [\Pi_{S}(\mathbf{P},t_s)] [\Pi_{S}(\mathbf{P},0)]^\dagger\right\rangle.
\end{align}
Here, $\Pi_{S}=\pi_{S}, K_S$ denote the pion and kaon interpolating operators, respectively, which can be written as follows
\begin{align}
\begin{split}
\pi_S(\mathbf{P},t)=&\sum_{\mathbf{x}}\bar{d}_S(\mathbf{x},t)\gamma_5u_S(\mathbf{x},t)e^{-i\mathbf{P}\cdot\mathbf{x}},\\
K_S(\mathbf{P},t)=&\sum_{\mathbf{x}}\bar{s}_S(\mathbf{x},t)\gamma_5u_S(\mathbf{x},t)e^{-i\mathbf{P}\cdot\mathbf{x}}.\\
\end{split}
\end{align}
These interpolating operators are constructed from Gaussian-smeared quark sources (sinks) in
Coulomb gauge 
\cite{Izubuchi:2019lyk}, which are also boosted with momentum  $\mathbf{k}^i$ ($\mathbf{k}^f$)~\cite{Bali:2016lva, Izubuchi:2019lyk}. Hence, we use the subscript $S$
in the above equations.
The Gaussian radii of the light and strange quarks used in this work are $r_l^G$ = 0.59 fm and $r_s^G$ = 0.83 fm for the $a$ = 0.076 fm lattice, and $r_l^G$ = 0.59 fm and $r_s^G$ = 0.86 fm for the $a$ = 0.04 fm lattice.

The three-point functions for EMFF can be written as,
\begin{align}
\begin{split}
    &C_{\rm 3pt}(\mathbf{P}^f,\mathbf{P}^i;\tau,t_s)\\
    =&\left\langle [\Pi_{S^f}(\mathbf{P}^f,t_s)] O_{\Gamma}(\mathbf{q},\tau) [\Pi_{S^i}(\mathbf{P}^i,0)]^\dagger\right\rangle,
\end{split}
\end{align}
with $\mathbf{P}^i=\mathbf{P}^f-\mathbf{q}$. 
Here, the electro-magnetic current is
$O_{\Gamma}=\frac{2}{3}\bar{u}\gamma_\mu u-\frac{1}{3}\bar{d}\gamma_\mu d$ and $\frac{2}{3}\bar{u}\gamma_\mu u-\frac{1}{3}\bar{s}\gamma_\mu s$ for pion and kaon, respectively. The time component of the vector current $\gamma_\mu=\gamma_0$ is used. 
With degenerate light quark masses, there are no disconnected diagrams for the pion, while their contribution to the kaon, 
which is expected to be small, is neglected in this work. To achieve high momentum transfer $Q^2=-(p^f-p^i)^2$ 
as the main target of this work, we make use of the Breit frame 
with $\mathbf{P}^f=(0,0, -P_3)$ and $\mathbf{q}=(0,0,-2P_3)$, when calculating the three-point functions. Here $P_3 = 2\pi n_3/(L_s a)$ with $n_3$ being an integer.
To optimize the signal, we use the same quark boost parameter $\zeta$ for the hadron states moving 
back-to-back, meaning that for the  quark momentum boost,
we choose $\mathbf{k}^f$ = $-\mathbf{k}^i$ to ensure $\zeta=\mathbf{k}^i/\mathbf{P}^i=\mathbf{k}^f/\mathbf{P}^f$
~\cite{Gao:2022iex, Gao:2020ito}. Given that the hadron states with slight momentum variation share the same propagator~\cite{Gao:2022iex}, 
it is possible to reliably calculate the three-point functions at multiple momentum transfers with small deviations from the Breit frame, requiring minimal additional computational costs.
For instance, in the case of the kaon, apart from the Breit frame scenario with $\mathbf{P}^f=(0,0,-2.42)$ GeV and $\mathbf{P}^i=(0,0,2.42)$ GeV, we can also
consider the non-Breit frame setup with $\mathbf{P}^f=(0,0,-2.42)$ GeV and $\mathbf{P}^i=(0,0,2.91)$ GeV, which allows us to reach $Q^2$ up to $28~\mathrm{GeV}^2$.
In this study, we used 350 gauge configurations
for the $a=0.076$ fm lattice and 280 gauge configurations for the $a=0.04$ fm lattice.
The number of AMA samples ranged from 32 to 256, depending on the lattice spacing and the momentum considered, with more
samples used for larger momenta. More detailed information on our choice of momenta and the number of AMA samples can
be found in the Supplemental Material.

\begin{figure}
\centering
    \includegraphics[width=0.42\textwidth]{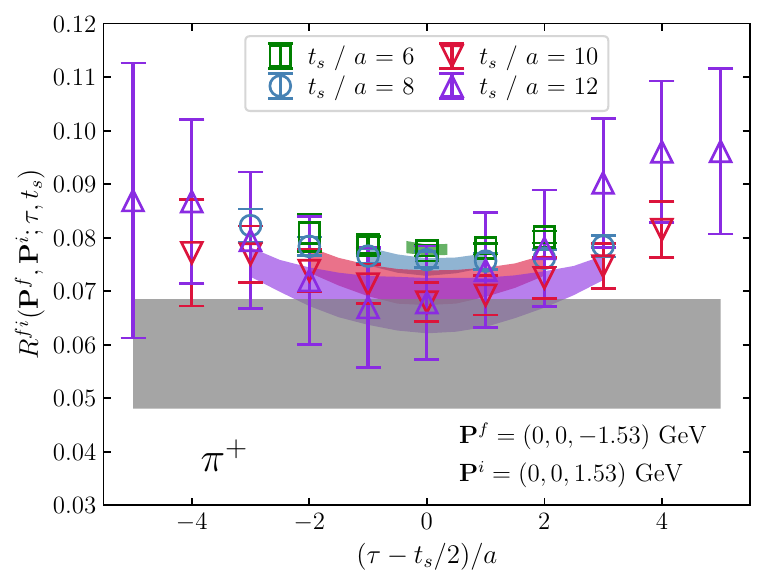}
    \includegraphics[width=0.42\textwidth]{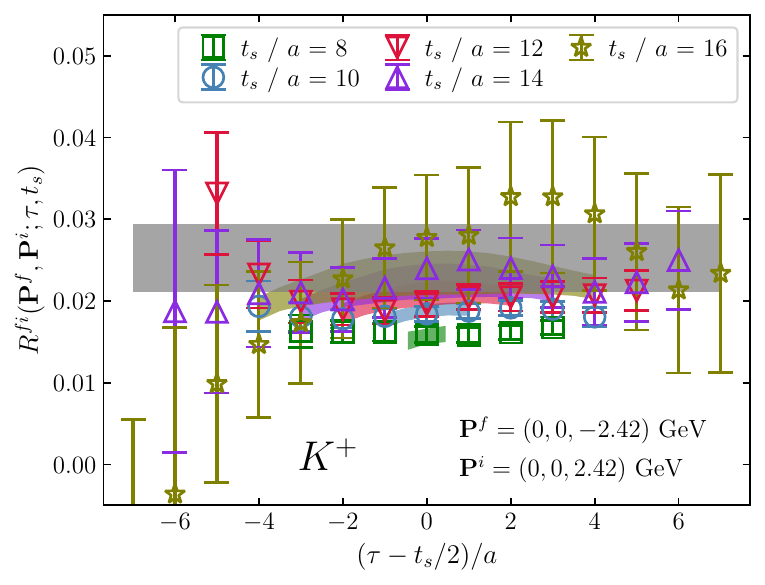}
	\caption{Upper panel: ratios $R^{fi}(t_s,\tau)$ for the pion with $Q^2=9.4~\rm GeV^2$ from the $a=0.076$ fm lattice. Lower panel: ratios $R^{fi}(t_s,\tau)$ for the kaon with $Q^2=23.4~\rm GeV^2$ from the $a=0.04$ fm lattice. The colored bands denote the two-state fit results of the corresponding colored lattice data, and the grey bands display the bare matrix elements of the ground state.\label{fig:fitcmp}}
\end{figure}

To take advantage of the correlation between the two-point and three-point functions, we construct the ratio
\begin{align}\label{eq:ratio}
	\begin{split}
	&R^{fi}(\mathbf{P}^f,\mathbf{P}^i;\tau,t_s) \equiv \frac{2\sqrt{E_0^f E_0^i}}{E_0^f+E_0^i} \frac{C_{\rm 3pt}(\mathbf{P}^f, \mathbf{P}^i; \tau,t_s)}{C_{\rm 2pt}(\mathbf{P}^f, t_s)}\\
	&\times \left[ \frac{C_{\rm 2pt}(\mathbf{P}^i, t_s - \tau)C_{\rm 2pt}(\mathbf{P}^f, \tau)C_{\rm 2pt}(\mathbf{P}^f, t_s)}{C_{\rm 2pt}(\mathbf{P}^f, t_s - \tau)C_{\rm 2pt}(\mathbf{P}^i, \tau)C_{\rm 2pt}(\mathbf{P}^i, t_s)} \right]^{1/2}. \\
	\end{split} 
\end{align}
In the $t_s\rightarrow\infty$ limit, this ratio approaches the ground-state bare matrix 
elements $R^{fi}(\mathbf{P}^f,\mathbf{P}^i;\tau\rightarrow \infty,t_s\rightarrow \infty) = F^B_M(Q^2)$ ($M=\pi^+$ or $K^+$). Taking the results of the energy levels and overlap amplitudes from the analysis of the two-point functions, 
we use the two-state fit on the lattice data to extract the bare matrix elements from the ratios. 
Details of the analysis can be found in the Supplemental Material. We show examples in the Breit frame of the ratio $R^{fi}$ in \fig{fitcmp} for the pion at $Q^2=9.4~\rm GeV^2$ on the $a=0.076$ fm lattice (upper panel) 
and the kaon at $Q^2=23.4~\rm GeV^2$ on the $a=0.04$ fm lattice (lower panel). 
Despite the very large momentum transfer, reasonable signals can be obtained by using the optimized boosted sources 
and large statistics. The bands in the figure show the results of the two-state fit with the corresponding statistical
errors obtained from the bootstrap analysis.
As one can see from the figure, the fit results describe the data very well. 
The grey bands display the results at the limits of infinite source-sink separation, giving the bare matrix elements of the pion and kaon ground states, that is, the bare form factors. These bare matrix elements $F^B_M(Q^2)$ are then converted to physical form factor values $F_M(Q^2)=Z_V F^B_M(Q^2)$ using the vector current renormalization factors $Z_V$.
We use the values of $Z_V$ calculated in our previous studies~\cite{Gao:2020ito, Gao:2021xsm} employing the same lattice setups,
specifically $1/Z_V$ = 1.048(2) and 1.024(1) for the $a = 0.076$ fm and 0.04 fm lattices, respectively.

\begin{figure}
\centering
    \includegraphics[width=0.45\textwidth]{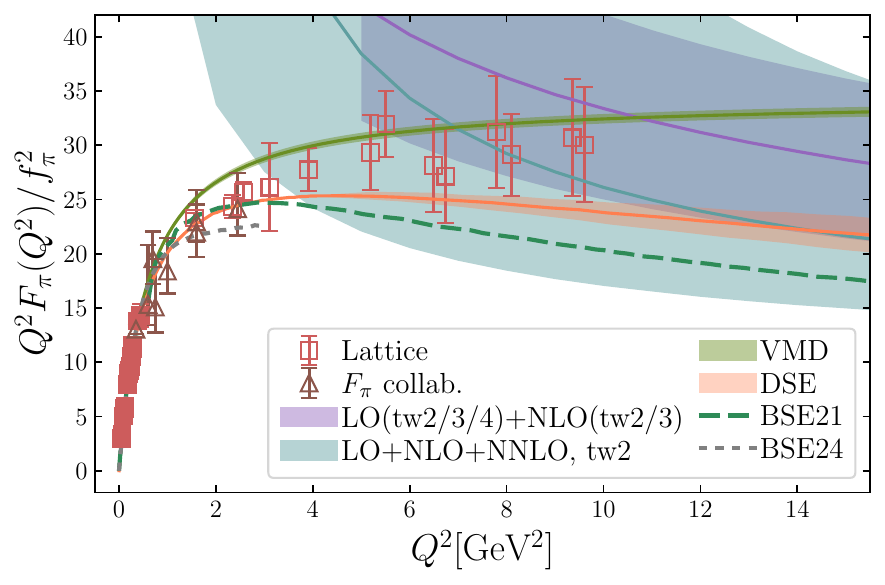}
    \includegraphics[width=0.45\textwidth]{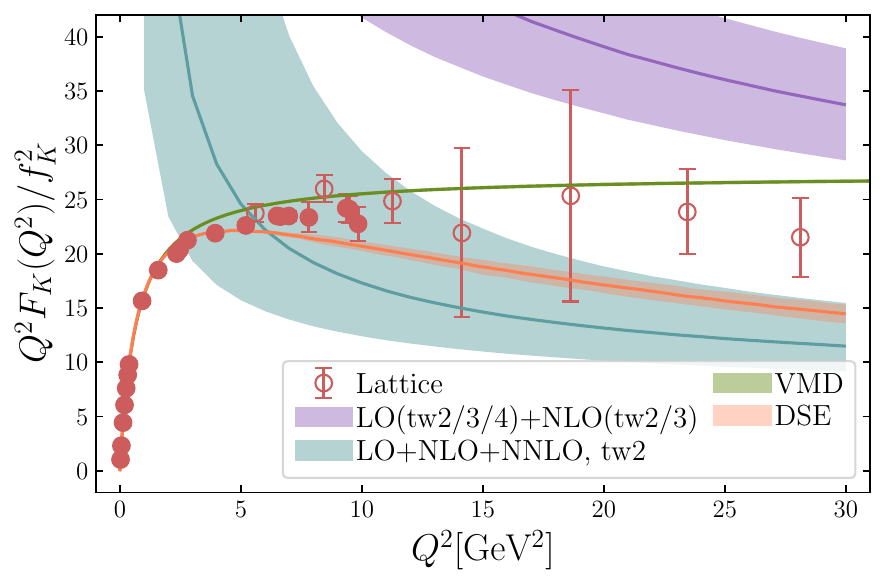} 
    \caption{The renormalized EMFF of the pion (upper panel) and kaon (lower panel) are shown as $Q^2F_{M}(Q^2)/f^2_M$.
    For the case of pion, we include the low $Q^2$ results (filled square symbols) from \refcite{Gao:2021xsm} that
    uses the same lattice setup, as well as the results from $F_\pi$ 
    collaboration~\cite{JeffersonLab:2008jve} extracted from the experimental data (open triangle symbols). 
    For the case of kaon, we show the results from two different lattices with 
    $a=0.076$ fm (filled circle symbols) and $a=0.04$ fm (open circle symbols). 
    The blue bands represent the twist two (tw2) pQCD results using the collinear
    factorization at NNLO. The purple bands denote the pQCD results
    obtained within the $k_T$ factorization theorem, which includes higher-twist contributions from \refcites{Cheng:2019ruz, Chai:2022srx, New_kaon_pQCD};
    see text. The width of the band presents the perturbative uncertainty.
    In the case of the NNLO twist two results, it corresponds to
    the scale variation from $\mu=Q/2$ to $\mu=2 Q$ and also includes
    the uncertainties of the conformal moments of the pion and kaon DA;
    see text.
    The green bands show the predictions with the VMD model by the fit on the lattice data at the low $Q^2$ region; see text. The dashed lines display the predictions from the DSE~\cite{Yao:2024drm}, BSE21~\cite{Ydrefors:2021dwa} and BSE24~\cite{Jia:2024dfl}; see text.
    \label{fig:EMFFLat}}
\end{figure}

%%%%%%%%%%%%%%%%%%%%%%%%%%%%%%%%%%%%%%%%%%%%%%%%%%%%%%%%%%%%%%%%%%%%%%
\textit{Form factors from low to high $Q^2$.} 
Our results of the renormalized EMFF for the pion and the kaon are shown in \fig{EMFFLat} as the combination  $Q^2F_{M}(Q^2)/f^2_M$, where $f_M$ is the meson decay constant for $M=\pi^+$ or $K^+$, the reason for this combination will become clear below.
We use the following values of the pion and kaon decay constant from FLAG21~\cite{Aoki:2021kgd, RBC:2014ntl, Follana:2007uv, MILC:2010hzw}: 
$f_{\pi} = 130.2$ MeV and $f_K=155.7$ MeV. 
For the pion, we also show the lattice results at low $Q^2$ (BNL21) from \refcite{Gao:2021xsm} 
obtained using the same lattice setup with $a$ = 0.076 fm. Furthermore, in the upper panel of
the figure, we compare the lattice results on $F_{\pi^+}(Q^2)$ to the 
experimental results from $F_\pi$ collaboration~\cite{JeffersonLab:2008jve}. 
Impressively, the lattice determination exhibits excellent agreement with the experimental data.
This agreement is reassuring for the model-based extraction of meson EMFF from experimental data, 
further bolstering the anticipation for their future experimental determination at JLAB~\cite{Dudek:2012vr, Arrington:2021alx}, EIC~\cite{Arrington:2021biu}, and EicC~\cite{Anderle:2021wcy}.
For the kaon, we have results from two different lattice ensembles with $a=0.076$ fm (filled circle symbols) 
and $a=0.04$ fm (open circle symbols), which appear to be consistent at the overlap range. This suggests that
the lattice artifacts are small compared to the statistical errors. 
Overall, one can observe that both $Q^2F_{\pi^+}(Q^2)$ and $Q^2F_{K^+}(Q^2)$ 
exhibit a rapid rise at low $Q^2$, 
transitioning into a plateau-like region at high $Q^2\gtrsim5\rm~ GeV^2$ within the errors. 

At high $Q^2$, the asymptotic nature of QCD allows the factorization of the electromagnetic form factors 
into the convolutions of the meson DA and a hard-scattering kernel,
as has been pointed out in the past~\cite{Efremov:1979qk, Farrar:1979aw, Lepage:1980fj}. Therefore, our lattice QCD results
can be utilized to test this factorization.
At the leading twist, the collinear factorization formula of the form factor reads
\begin{align}\label{eq:factorization}
\begin{split}
    F_M(Q^2)& =\int^1_0\int^1_0dxdy\ \Phi^*_M(y,\mu^2_F) \\
    &\times T_H(x,y,Q^2,\mu^2_R,\mu^2_F)\Phi_M(x,\mu^2_F),
\end{split}
\end{align}
where $T_H$ is the hard-process kernel calculated in perturbative QCD (pQCD). 
The hard kernel depends on the momentum transfer $Q^2$, the factorization scale $\mu_F$, as well as the renormalization scale $\mu_R$ at a fixed order of perturbation theory. It has been known up to the next-to-leading order (NLO)~\cite{Field:1981wx, Dittes:1981aw, Sarmadi:1982yg, Braaten:1987yy} for some time. Very recently, the NNLO
correction has become available~\cite{Chen:2023byr}.
%\cite{Nandi:2007qx,Li:2010nn}.
The nonperturbative physics is encoded in the meson DA $\Phi_M(x,\mu^2_F)$. Its dependence on $\mu_F$ comes from its anomalous dimension, which is compensated by the $\mu_F$-dependence of the hard kernel in the factorization formula above. 
The limit of very large $Q^2$ also implies that $\mu_F \rightarrow \infty$, and in this limit, one can use the asymptotic limit of DA, given by $\Phi_M^{\rm as}=3 f_M x (1-x)/\sqrt{2 N_c}$ ($N_c=3$ for QCD). 
Therefore, at asymptotically large $Q^2$, where the leading order (LO) result for $T_H$ is justified, we have 
$F_M(Q^2 \rightarrow \infty) = 8 \pi \alpha_s(Q^2) f_M^2/Q^2$. This means that $Q^2 F_M(Q^2)/f_M^2$ should be 
approximately constant and small at sufficiently large $Q^2$, explaining our normalization process. This LO asymptotic result gives a very small EMFF in comparison
to the experimental results. For example, for the largest $Q^2=2.45~{\rm GeV}^2$ accessible
experimentally,
the LO asymptotic result predicts $Q^2 F_{\pi^+}(Q^2)/f_{\pi}^2 \simeq 8.6$, which is almost three times smaller than the experimental value, as shown in \fig{EMFFLat}.

The pion and kaon EMFF have also been calculated using the $k_T$ factorization approach~\cite{Nandi:2007qx, Li:2010nn, Raha:2008ve, Cheng:2019ruz, Chai:2022srx}.
For $Q\gg k_T$, the pion and kaon EMFF can also be related to the meson DA in this approach.
Furthermore, it has long been suggested that higher-twist contributions to the EMFF factorization
could be numerically large even for $Q^2<100$ GeV$^2$ 
despite being formally suppressed~\cite{Geshkenbein:1982zs, Cao:1997st, Braun:1999uj,Bijnens:2002mg, Wei:2003fm, Huang:2004su, Raha:2008ve, Cheng:2019ruz, Chai:2022srx}.
The state-of-the-art studies that consider higher-twist contributions are performed within the $k_T$ factorization approach and include all two- and
three-particle (parton) contributions up to twist four~\cite{Cheng:2019ruz, Chai:2022srx}. 
We note, however, that it has been known since long that there are infrared sensitive double logs
in the hard kernel at twist three and beyond \cite{Geshkenbein:1982zs,Braun:1999uj}, and it is not
completely clear that these are properly treated in the $k_T$ factorization.
In \fig{EMFFLat}, we compare our lattice results with the state-of-the-art perturbative calculations, including the higher-twist contributions
under the $k_T$ factorization framework~\cite{Cheng:2019ruz, Chai:2022srx, New_kaon_pQCD}, 
as well as the pQCD predictions up to NNLO within the collinear factorization framework~\cite{Chen:2023byr}.
%For the comparison with the lattice QCD results
For the latter, we use the NNLO perturbative results written
in terms of Gegenbauer expansion of the pion and kaon DA. The DAs can be calculated from lattice QCD~\cite{RQCD:2019osh,Detmold:2021qln,LatticeParton:2022zqc,Gao:2022vyh}. In this work, we utilize results obtained from the same lattice setup using the $a$ = 0.076 fm gauge ensemble~\cite{Cloet:2024vbv}, where we derived up to the sixth conformal moment with minimal model dependence. We use the conformal moments at $\mu_F=2$ GeV:
$a_2=0.196(32),~a_4=0.085(26),~a_6=0.056(15)$ for the pion
and $a_2=0.114(20),~a_4=0.037(11),~a_6=0.019(5)$ for the kaon. We set $\mu_F=\mu_R=Q$ and evolve the conformal moments consistently using their anomalous dimensions up to 3 loops~\cite{Braun:2017cih}. We vary the scales between $Q/2$ and $2 Q$ to estimate the theoretical uncertainties. 

We see from \fig{EMFFLat} that for $Q^2>5~{\rm GeV}^2$ lattice QCD results for pion and kaon EMFF mostly agree
with the collinear NNLO pQCD results within the estimated errors. Note that the same values of DAs used to produce the EMFF also lead to prediction for pion transition form factors~\cite{Cloet:2024vbv} that are consistent with experimental data from the Belle collaboration~\cite{Belle:2012wwz}. For the first time, the perturbative factorization and universality of DAs are tested in two exclusive processes, thanks to the large momentum transfer achieved in this work. For the pion, we also find agreement with NLO results from the $k_T$ factorization that includes higher-twist contributions~\cite{Cheng:2019ruz, Chai:2022srx}.
However, for the kaon case, the $k_T$ factorization approach with higher-twist contributions overestimates the lattice QCD
results and also leads to much stronger $Q^2$-dependence of $Q^2 F_K(Q^2)$ compared to our lattice results. 

The fact that LO asymptotic pQCD prediction results in small EMFF at large $Q^2$ 
in comparison to the experimental results, motivated the development of 
partonic approaches that incorporate
some nonperturbative information. These approaches include those based on the Dyson-Schwinger
equation (DSE)~\cite{Gao:2017mmp, Yao:2024drm} and the Bethe-Salpeter equation (BSE)~\cite{Ydrefors:2021dwa, Jia:2024dfl}.
The former approach is related to the idea of dynamical mass generation in QCD.
We also compare our lattice results against these theoretical predictions in \fig{EMFFLat}.
For the pion EMFF, there is a reasonable agreement between the lattice QCD results
and the DSE calculations~\cite{Gao:2017mmp, Yao:2024drm}, whereas the BSE calculations~\cite{Ydrefors:2021dwa, Jia:2024dfl} fall below the lattice results for $Q^2\gtrsim3$ GeV$^2$.
In the case of the kaon EMFF, there is some tension between the DSE calculations and
the lattice results during the intermediate $Q^2$ range.

At low $Q^2$, the form factors could be understood by low-energy models 
such as the Vector Meson Dominance (VMD) model~\cite{OConnell:1995fwv, Klingl:1996by}. 
Based on the lowest-lying vector resonances, the
kaon form factors can be parameterized as~\cite{Klingl:1996by, Ivashyn:2006gf, Stamen:2022uqh},
\begin{align}
	F_{K^+}(Q^2)=\sum_v\frac{c_v}{1+Q^2/m_v^2},
\label{eq:VMD}
\end{align}
with $m_v$ being the mass of vector mesons. In this work, we take $v=\rho,\phi,\omega$ with 
their mass from PDG~\cite{Workman:2022ynf} and fit $c_v$ using the lattice results at low $Q^2\leq0.4\rm~GeV^2$. 
Using this parameterization, the charge radius of $K^+$ can be determined by the slope at $F_{K^+}(0)$,
which gives $\langle r^2_K\rangle = 0.360(2)$ fm$^2$. This value is consistent with a 
recent dispersive analysis of the experimental data that gives 
$\langle r^2_K\rangle = 0.359(3)$ fm$^2$~\cite{Stamen:2022uqh}. As for the pion, 
the form factors at low $Q^2$ can be parameterized by a single term in Eq.~(\ref{eq:VMD}) using the VMD form, which is a one-parameter fit as discussed in \refcite{Gao:2021xsm}. The fit results from the VMD model 
extended to a high $Q^2$ region are also shown in \fig{EMFFLat}.
Surprisingly, the VMD fits from the low $Q^2$ region give a fairly good description
of the lattice results for the pion and kaon EMFF, even for relatively large $Q^2$ values, up to $Q^2=5~{\rm GeV}^2$.

Since we have information on the pion and kaon form factors up to high $Q^2$, we can extract 
the charge or flavor distribution of hadrons in the impact 
parameter plane~\cite{Bhattacharya:2023ays} using a model-independent way. There is no need
to model the high $Q^2$ behavior of the form factors.
We find that the form factors as well as the impact parameter ($b_\perp$)
distributions of the $u$ quark inside  
the pion and the kaon are identical, contradicting to the prediction from NJL model~\cite{Hutauruk:2016sug}. The distribution
of the heavier $s$ anti-quark, on the other hand, is considerably narrower.
The details of these calculations are presented in the Supplemental Material.

%%%%%%%%%%%%%%%%%%%%%%%%%%%%%%%%%%%%%%%%%%%%%%%%%%%%%%%%%%%%%%%%%%%%%%
\textit{Summary.} In this work, we calculated the electromagnetic form factors 
of the pion and kaon with large $Q^2$ up to $10 ~\rm GeV^2$ and $28 ~\rm GeV^2$, respectively,
on the lattice for the first time. For the kaon EMFF, we employ two lattice ensembles and find that
the lattice artifacts are minor compared to the statistical uncertainties. Future stuties targeting precision at percent level should be more careful with the systematics including the excited state contamination and discretization effect.
Our results can serve as benchmark QCD predictions for model-based studies and 
for the forthcoming experimental measurements planned at JLab and the future EIC and EicC. 
We find that for $Q^2>5~{\rm GeV}^2$ the lattice results on the pion and kaon EMFF
agree with the leading-twist collinear factorization if the NNLO hard kernel is used together
with the most recent lattice QCD results on the DA. For smaller values of $Q^2$, our lattice
QCD results can be understood in terms of the VMD model, suggesting that the transition from
hadronic description to partonic description does not lead to abrupt changes in the $Q^2$-dependence
of the pion and kaon EMFF.
Finally, our results for the form factors provide a model-independent picture
of the spatial distribution in the transverse plane of the up and strange quarks inside the pion and the kaon.

\begin{acknowledgments}

We thank Vladimir M. Braun for his valuable communications. This material is based upon work supported by The U.S. Department of Energy, Office of Science, Office of Nuclear Physics through Contract No.~DE-SC0012704, Contract No.~DE-AC02-06CH11357, and within the frameworks of Scientific Discovery through Advanced Computing (SciDAC) award Fundamental Nuclear Physics at the Exascale and Beyond and the Topical Collaboration in Nuclear Theory 3D quark-gluon structure of hadrons: mass, spin, and tomography.
YZ is partially supported by the 2023 Physical Sciences and Engineering (PSE) Early Investigator Named Award program at Argonne National Laboratory.
SS is supported by the National Science Foundation under CAREER Award PHY-1847893.
HTD is supported by the NSFC under grant No.~12325508, and the National Key Research and Development Program of China under Grant No. 2022YFA1604900.

This research used awards of computer time provided by the INCITE program at Argonne Leadership Computing Facility, a DOE Office of Science User Facility operated under Contract DE-AC02-06CH11357, the ALCC program at the Oak Ridge Leadership Computing Facility, which is a DOE Office of Science User Facility supported under Contract DE-AC05-00OR22725, the National Energy Research
Scientific Computing Center, a DOE Office of Science User Facility supported by the Office of Science of the U.S. Department of Energy under Contract DE-AC02-05CH11231 using NERSC award NP-ERCAP0028137. Computations for this work were carried out in part on facilities of the USQCD Collaboration, which is funded by the Office of Science of the U.S. Department of Energy, and the Nuclear Science Computing Center at Central China Normal University ($\mathrm{NSC}^{3}$) and Wuhan Supercomputing Center.

\end{acknowledgments}

\bibliographystyle{apsrev4-2.bst}

\bibliography{ref}

\begin{thebibliography}{93}%
\makeatletter
\providecommand \@ifxundefined [1]{%
 \@ifx{#1\undefined}
}%
\providecommand \@ifnum [1]{%
 \ifnum #1\expandafter \@firstoftwo
 \else \expandafter \@secondoftwo
 \fi
}%
\providecommand \@ifx [1]{%
 \ifx #1\expandafter \@firstoftwo
 \else \expandafter \@secondoftwo
 \fi
}%
\providecommand \natexlab [1]{#1}%
\providecommand \enquote  [1]{``#1''}%
\providecommand \bibnamefont  [1]{#1}%
\providecommand \bibfnamefont [1]{#1}%
\providecommand \citenamefont [1]{#1}%
\providecommand \href@noop [0]{\@secondoftwo}%
\providecommand \href [0]{\begingroup \@sanitize@url \@href}%
\providecommand \@href[1]{\@@startlink{#1}\@@href}%
\providecommand \@@href[1]{\endgroup#1\@@endlink}%
\providecommand \@sanitize@url [0]{\catcode `\\12\catcode `\$12\catcode
  `\&12\catcode `\#12\catcode `\^12\catcode `\_12\catcode `\%12\relax}%
\providecommand \@@startlink[1]{}%
\providecommand \@@endlink[0]{}%
\providecommand \url  [0]{\begingroup\@sanitize@url \@url }%
\providecommand \@url [1]{\endgroup\@href {#1}{\urlprefix }}%
\providecommand \urlprefix  [0]{URL }%
\providecommand \Eprint [0]{\href }%
\providecommand \doibase [0]{https://doi.org/}%
\providecommand \selectlanguage [0]{\@gobble}%
\providecommand \bibinfo  [0]{\@secondoftwo}%
\providecommand \bibfield  [0]{\@secondoftwo}%
\providecommand \translation [1]{[#1]}%
\providecommand \BibitemOpen [0]{}%
\providecommand \bibitemStop [0]{}%
\providecommand \bibitemNoStop [0]{.\EOS\space}%
\providecommand \EOS [0]{\spacefactor3000\relax}%
\providecommand \BibitemShut  [1]{\csname bibitem#1\endcsname}%
\let\auto@bib@innerbib\@empty
%</preamble>
\bibitem [{\citenamefont {Epelbaum}\ \emph {et~al.}(2022)\citenamefont
  {Epelbaum}, \citenamefont {Gegelia}, \citenamefont {Lange}, \citenamefont
  {Mei\ss{}ner},\ and\ \citenamefont {Polyakov}}]{Epelbaum:2022fjc}%
  \BibitemOpen
  \bibfield  {author} {\bibinfo {author} {\bibfnamefont {E.}~\bibnamefont
  {Epelbaum}}, \bibinfo {author} {\bibfnamefont {J.}~\bibnamefont {Gegelia}},
  \bibinfo {author} {\bibfnamefont {N.}~\bibnamefont {Lange}}, \bibinfo
  {author} {\bibfnamefont {U.~G.}\ \bibnamefont {Mei\ss{}ner}},\ and\ \bibinfo
  {author} {\bibfnamefont {M.~V.}\ \bibnamefont {Polyakov}},\ }\href
  {https://doi.org/10.1103/PhysRevLett.129.012001} {\bibfield  {journal}
  {\bibinfo  {journal} {Phys. Rev. Lett.}\ }\textbf {\bibinfo {volume} {129}},\
  \bibinfo {pages} {012001} (\bibinfo {year} {2022})},\ \Eprint
  {https://arxiv.org/abs/2201.02565} {arXiv:2201.02565 [hep-ph]} \BibitemShut
  {NoStop}%
\bibitem [{\citenamefont {Hofstadter}\ \emph {et~al.}(1958)\citenamefont
  {Hofstadter}, \citenamefont {Bumiller},\ and\ \citenamefont
  {Yearian}}]{RevModPhys.30.482}%
  \BibitemOpen
  \bibfield  {author} {\bibinfo {author} {\bibfnamefont {R.}~\bibnamefont
  {Hofstadter}}, \bibinfo {author} {\bibfnamefont {F.}~\bibnamefont
  {Bumiller}},\ and\ \bibinfo {author} {\bibfnamefont {M.~R.}\ \bibnamefont
  {Yearian}},\ }\href {https://doi.org/10.1103/RevModPhys.30.482} {\bibfield
  {journal} {\bibinfo  {journal} {Rev. Mod. Phys.}\ }\textbf {\bibinfo {volume}
  {30}},\ \bibinfo {pages} {482} (\bibinfo {year} {1958})}\BibitemShut
  {NoStop}%
\bibitem [{\citenamefont {Cui}\ \emph {et~al.}(2021)\citenamefont {Cui},
  \citenamefont {Ding}, \citenamefont {Gao}, \citenamefont {Raya},
  \citenamefont {Binosi}, \citenamefont {Chang}, \citenamefont {Roberts},
  \citenamefont {Rodriguez-Quintero},\ and\ \citenamefont
  {Schmidt}}]{Cui:2020dlm}%
  \BibitemOpen
  \bibfield  {author} {\bibinfo {author} {\bibfnamefont {Z.-F.}\ \bibnamefont
  {Cui}}, \bibinfo {author} {\bibfnamefont {M.}~\bibnamefont {Ding}}, \bibinfo
  {author} {\bibfnamefont {F.}~\bibnamefont {Gao}}, \bibinfo {author}
  {\bibfnamefont {K.}~\bibnamefont {Raya}}, \bibinfo {author} {\bibfnamefont
  {D.}~\bibnamefont {Binosi}}, \bibinfo {author} {\bibfnamefont
  {L.}~\bibnamefont {Chang}}, \bibinfo {author} {\bibfnamefont {C.~D.}\
  \bibnamefont {Roberts}}, \bibinfo {author} {\bibfnamefont {J.}~\bibnamefont
  {Rodriguez-Quintero}},\ and\ \bibinfo {author} {\bibfnamefont {S.~M.}\
  \bibnamefont {Schmidt}},\ }\href
  {https://doi.org/10.1140/epja/s10050-020-00318-2} {\bibfield  {journal}
  {\bibinfo  {journal} {Eur. Phys. J. A}\ }\textbf {\bibinfo {volume} {57}},\
  \bibinfo {pages} {5} (\bibinfo {year} {2021})},\ \Eprint
  {https://arxiv.org/abs/2006.14075} {arXiv:2006.14075 [hep-ph]} \BibitemShut
  {NoStop}%
\bibitem [{\citenamefont {Roberts}\ \emph {et~al.}(2021)\citenamefont
  {Roberts}, \citenamefont {Richards}, \citenamefont {Horn},\ and\
  \citenamefont {Chang}}]{Roberts:2021nhw}%
  \BibitemOpen
  \bibfield  {author} {\bibinfo {author} {\bibfnamefont {C.~D.}\ \bibnamefont
  {Roberts}}, \bibinfo {author} {\bibfnamefont {D.~G.}\ \bibnamefont
  {Richards}}, \bibinfo {author} {\bibfnamefont {T.}~\bibnamefont {Horn}},\
  and\ \bibinfo {author} {\bibfnamefont {L.}~\bibnamefont {Chang}},\ }\href
  {https://doi.org/10.1016/j.ppnp.2021.103883} {\bibfield  {journal} {\bibinfo
  {journal} {Prog. Part. Nucl. Phys.}\ }\textbf {\bibinfo {volume} {120}},\
  \bibinfo {pages} {103883} (\bibinfo {year} {2021})},\ \Eprint
  {https://arxiv.org/abs/2102.01765} {arXiv:2102.01765 [hep-ph]} \BibitemShut
  {NoStop}%
\bibitem [{\citenamefont {Alberg}\ and\ \citenamefont
  {Miller}(2024)}]{Alberg:2024svo}%
  \BibitemOpen
  \bibfield  {author} {\bibinfo {author} {\bibfnamefont {M.}~\bibnamefont
  {Alberg}}\ and\ \bibinfo {author} {\bibfnamefont {G.~A.}\ \bibnamefont
  {Miller}},\ }\href@noop {} {\  (\bibinfo {year} {2024})},\ \Eprint
  {https://arxiv.org/abs/2403.03356} {arXiv:2403.03356 [hep-ph]} \BibitemShut
  {NoStop}%
\bibitem [{\citenamefont {Brambilla}\ \emph {et~al.}(2014)\citenamefont
  {Brambilla} \emph {et~al.}}]{Brambilla:2014jmp}%
  \BibitemOpen
  \bibfield  {author} {\bibinfo {author} {\bibfnamefont {N.}~\bibnamefont
  {Brambilla}} \emph {et~al.},\ }\href
  {https://doi.org/10.1140/epjc/s10052-014-2981-5} {\bibfield  {journal}
  {\bibinfo  {journal} {Eur. Phys. J. C}\ }\textbf {\bibinfo {volume} {74}},\
  \bibinfo {pages} {2981} (\bibinfo {year} {2014})},\ \Eprint
  {https://arxiv.org/abs/1404.3723} {arXiv:1404.3723 [hep-ph]} \BibitemShut
  {NoStop}%
\bibitem [{\citenamefont {Huber}\ \emph
  {et~al.}(2008{\natexlab{a}})\citenamefont {Huber} \emph
  {et~al.}}]{Huber:2008id}%
  \BibitemOpen
  \bibfield  {author} {\bibinfo {author} {\bibfnamefont {G.~M.}\ \bibnamefont
  {Huber}} \emph {et~al.} (\bibinfo {collaboration} {Jefferson Lab}),\ }\href
  {https://doi.org/10.1103/PhysRevC.78.045203} {\bibfield  {journal} {\bibinfo
  {journal} {Phys. Rev. C}\ }\textbf {\bibinfo {volume} {78}},\ \bibinfo
  {pages} {045203} (\bibinfo {year} {2008}{\natexlab{a}})},\ \Eprint
  {https://arxiv.org/abs/0809.3052} {arXiv:0809.3052 [nucl-ex]} \BibitemShut
  {NoStop}%
\bibitem [{\citenamefont {Farrar}\ and\ \citenamefont
  {Jackson}(1979)}]{Farrar:1979aw}%
  \BibitemOpen
  \bibfield  {author} {\bibinfo {author} {\bibfnamefont {G.~R.}\ \bibnamefont
  {Farrar}}\ and\ \bibinfo {author} {\bibfnamefont {D.~R.}\ \bibnamefont
  {Jackson}},\ }\href {https://doi.org/10.1103/PhysRevLett.43.246} {\bibfield
  {journal} {\bibinfo  {journal} {Phys. Rev. Lett.}\ }\textbf {\bibinfo
  {volume} {43}},\ \bibinfo {pages} {246} (\bibinfo {year} {1979})}\BibitemShut
  {NoStop}%
\bibitem [{\citenamefont {Lepage}\ and\ \citenamefont
  {Brodsky}(1980)}]{Lepage:1980fj}%
  \BibitemOpen
  \bibfield  {author} {\bibinfo {author} {\bibfnamefont {G.~P.}\ \bibnamefont
  {Lepage}}\ and\ \bibinfo {author} {\bibfnamefont {S.~J.}\ \bibnamefont
  {Brodsky}},\ }\href {https://doi.org/10.1103/PhysRevD.22.2157} {\bibfield
  {journal} {\bibinfo  {journal} {Phys. Rev. D}\ }\textbf {\bibinfo {volume}
  {22}},\ \bibinfo {pages} {2157} (\bibinfo {year} {1980})}\BibitemShut
  {NoStop}%
\bibitem [{\citenamefont {Aubert}\ \emph {et~al.}(2009)\citenamefont {Aubert}
  \emph {et~al.}}]{BaBar:2009rrj}%
  \BibitemOpen
  \bibfield  {author} {\bibinfo {author} {\bibfnamefont {B.}~\bibnamefont
  {Aubert}} \emph {et~al.} (\bibinfo {collaboration} {BaBar}),\ }\href
  {https://doi.org/10.1103/PhysRevD.80.052002} {\bibfield  {journal} {\bibinfo
  {journal} {Phys. Rev. D}\ }\textbf {\bibinfo {volume} {80}},\ \bibinfo
  {pages} {052002} (\bibinfo {year} {2009})},\ \Eprint
  {https://arxiv.org/abs/0905.4778} {arXiv:0905.4778 [hep-ex]} \BibitemShut
  {NoStop}%
\bibitem [{\citenamefont {Uehara}\ \emph {et~al.}(2012)\citenamefont {Uehara}
  \emph {et~al.}}]{Belle:2012wwz}%
  \BibitemOpen
  \bibfield  {author} {\bibinfo {author} {\bibfnamefont {S.}~\bibnamefont
  {Uehara}} \emph {et~al.} (\bibinfo {collaboration} {Belle}),\ }\href
  {https://doi.org/10.1103/PhysRevD.86.092007} {\bibfield  {journal} {\bibinfo
  {journal} {Phys. Rev. D}\ }\textbf {\bibinfo {volume} {86}},\ \bibinfo
  {pages} {092007} (\bibinfo {year} {2012})},\ \Eprint
  {https://arxiv.org/abs/1205.3249} {arXiv:1205.3249 [hep-ex]} \BibitemShut
  {NoStop}%
\bibitem [{\citenamefont {Braun}\ \emph {et~al.}(2021)\citenamefont {Braun},
  \citenamefont {Manashov}, \citenamefont {Moch},\ and\ \citenamefont
  {Schoenleber}}]{Braun:2021grd}%
  \BibitemOpen
  \bibfield  {author} {\bibinfo {author} {\bibfnamefont {V.~M.}\ \bibnamefont
  {Braun}}, \bibinfo {author} {\bibfnamefont {A.~N.}\ \bibnamefont {Manashov}},
  \bibinfo {author} {\bibfnamefont {S.}~\bibnamefont {Moch}},\ and\ \bibinfo
  {author} {\bibfnamefont {J.}~\bibnamefont {Schoenleber}},\ }\href
  {https://doi.org/10.1103/PhysRevD.104.094007} {\bibfield  {journal} {\bibinfo
   {journal} {Phys. Rev. D}\ }\textbf {\bibinfo {volume} {104}},\ \bibinfo
  {pages} {094007} (\bibinfo {year} {2021})},\ \Eprint
  {https://arxiv.org/abs/2106.01437} {arXiv:2106.01437 [hep-ph]} \BibitemShut
  {NoStop}%
\bibitem [{\citenamefont {Gao}\ \emph {et~al.}(2022{\natexlab{a}})\citenamefont
  {Gao}, \citenamefont {Huber}, \citenamefont {Ji},\ and\ \citenamefont
  {Wang}}]{Gao:2021iqq}%
  \BibitemOpen
  \bibfield  {author} {\bibinfo {author} {\bibfnamefont {J.}~\bibnamefont
  {Gao}}, \bibinfo {author} {\bibfnamefont {T.}~\bibnamefont {Huber}}, \bibinfo
  {author} {\bibfnamefont {Y.}~\bibnamefont {Ji}},\ and\ \bibinfo {author}
  {\bibfnamefont {Y.-M.}\ \bibnamefont {Wang}},\ }\href
  {https://doi.org/10.1103/PhysRevLett.128.062003} {\bibfield  {journal}
  {\bibinfo  {journal} {Phys. Rev. Lett.}\ }\textbf {\bibinfo {volume} {128}},\
  \bibinfo {pages} {062003} (\bibinfo {year} {2022}{\natexlab{a}})},\ \Eprint
  {https://arxiv.org/abs/2106.01390} {arXiv:2106.01390 [hep-ph]} \BibitemShut
  {NoStop}%
\bibitem [{\citenamefont {Dally}\ \emph {et~al.}(1981)\citenamefont {Dally}
  \emph {et~al.}}]{Dally:1981ur}%
  \BibitemOpen
  \bibfield  {author} {\bibinfo {author} {\bibfnamefont {E.~B.}\ \bibnamefont
  {Dally}} \emph {et~al.},\ }\href {https://doi.org/10.1103/PhysRevD.24.1718}
  {\bibfield  {journal} {\bibinfo  {journal} {Phys. Rev. D}\ }\textbf {\bibinfo
  {volume} {24}},\ \bibinfo {pages} {1718} (\bibinfo {year}
  {1981})}\BibitemShut {NoStop}%
\bibitem [{\citenamefont {Dally}\ \emph {et~al.}(1982)\citenamefont {Dally}
  \emph {et~al.}}]{Dally:1982zk}%
  \BibitemOpen
  \bibfield  {author} {\bibinfo {author} {\bibfnamefont {E.~B.}\ \bibnamefont
  {Dally}} \emph {et~al.},\ }\href {https://doi.org/10.1103/PhysRevLett.48.375}
  {\bibfield  {journal} {\bibinfo  {journal} {Phys. Rev. Lett.}\ }\textbf
  {\bibinfo {volume} {48}},\ \bibinfo {pages} {375} (\bibinfo {year}
  {1982})}\BibitemShut {NoStop}%
\bibitem [{\citenamefont {Amendolia}\ \emph {et~al.}(1984)\citenamefont
  {Amendolia} \emph {et~al.}}]{Amendolia:1984nz}%
  \BibitemOpen
  \bibfield  {author} {\bibinfo {author} {\bibfnamefont {S.~R.}\ \bibnamefont
  {Amendolia}} \emph {et~al.},\ }\href
  {https://doi.org/10.1016/0370-2693(84)90655-5} {\bibfield  {journal}
  {\bibinfo  {journal} {Phys. Lett. B}\ }\textbf {\bibinfo {volume} {146}},\
  \bibinfo {pages} {116} (\bibinfo {year} {1984})}\BibitemShut {NoStop}%
\bibitem [{\citenamefont {Amendolia}\ \emph
  {et~al.}(1986{\natexlab{a}})\citenamefont {Amendolia} \emph
  {et~al.}}]{Amendolia:1986wj}%
  \BibitemOpen
  \bibfield  {author} {\bibinfo {author} {\bibfnamefont {S.~R.}\ \bibnamefont
  {Amendolia}} \emph {et~al.} (\bibinfo {collaboration} {NA7}),\ }\href
  {https://doi.org/10.1016/0550-3213(86)90437-2} {\bibfield  {journal}
  {\bibinfo  {journal} {Nucl. Phys. B}\ }\textbf {\bibinfo {volume} {277}},\
  \bibinfo {pages} {168} (\bibinfo {year} {1986}{\natexlab{a}})}\BibitemShut
  {NoStop}%
\bibitem [{\citenamefont {Brauel}\ \emph {et~al.}(1979)\citenamefont {Brauel},
  \citenamefont {Canzler}, \citenamefont {Cords}, \citenamefont {Felst},
  \citenamefont {Grindhammer}, \citenamefont {Helm}, \citenamefont {Kollmann},
  \citenamefont {Krehbiel},\ and\ \citenamefont {Schadlich}}]{Brauel:1979zk}%
  \BibitemOpen
  \bibfield  {author} {\bibinfo {author} {\bibfnamefont {P.}~\bibnamefont
  {Brauel}}, \bibinfo {author} {\bibfnamefont {T.}~\bibnamefont {Canzler}},
  \bibinfo {author} {\bibfnamefont {D.}~\bibnamefont {Cords}}, \bibinfo
  {author} {\bibfnamefont {R.}~\bibnamefont {Felst}}, \bibinfo {author}
  {\bibfnamefont {G.}~\bibnamefont {Grindhammer}}, \bibinfo {author}
  {\bibfnamefont {M.}~\bibnamefont {Helm}}, \bibinfo {author} {\bibfnamefont
  {W.~D.}\ \bibnamefont {Kollmann}}, \bibinfo {author} {\bibfnamefont
  {H.}~\bibnamefont {Krehbiel}},\ and\ \bibinfo {author} {\bibfnamefont
  {M.}~\bibnamefont {Schadlich}},\ }\href {https://doi.org/10.1007/BF01443698}
  {\bibfield  {journal} {\bibinfo  {journal} {Z. Phys. C}\ }\textbf {\bibinfo
  {volume} {3}},\ \bibinfo {pages} {101} (\bibinfo {year} {1979})}\BibitemShut
  {NoStop}%
\bibitem [{\citenamefont {Dally}\ \emph {et~al.}(1980)\citenamefont {Dally}
  \emph {et~al.}}]{Dally:1980dj}%
  \BibitemOpen
  \bibfield  {author} {\bibinfo {author} {\bibfnamefont {E.~B.}\ \bibnamefont
  {Dally}} \emph {et~al.},\ }\href {https://doi.org/10.1103/PhysRevLett.45.232}
  {\bibfield  {journal} {\bibinfo  {journal} {Phys. Rev. Lett.}\ }\textbf
  {\bibinfo {volume} {45}},\ \bibinfo {pages} {232} (\bibinfo {year}
  {1980})}\BibitemShut {NoStop}%
\bibitem [{\citenamefont {Amendolia}\ \emph
  {et~al.}(1986{\natexlab{b}})\citenamefont {Amendolia} \emph
  {et~al.}}]{Amendolia:1986ui}%
  \BibitemOpen
  \bibfield  {author} {\bibinfo {author} {\bibfnamefont {S.~R.}\ \bibnamefont
  {Amendolia}} \emph {et~al.},\ }\href
  {https://doi.org/10.1016/0370-2693(86)91407-3} {\bibfield  {journal}
  {\bibinfo  {journal} {Phys. Lett. B}\ }\textbf {\bibinfo {volume} {178}},\
  \bibinfo {pages} {435} (\bibinfo {year} {1986}{\natexlab{b}})}\BibitemShut
  {NoStop}%
\bibitem [{\citenamefont {Carmignotto}\ \emph {et~al.}(2018)\citenamefont
  {Carmignotto} \emph {et~al.}}]{Carmignotto:2018uqj}%
  \BibitemOpen
  \bibfield  {author} {\bibinfo {author} {\bibfnamefont {M.}~\bibnamefont
  {Carmignotto}} \emph {et~al.},\ }\href
  {https://doi.org/10.1103/PhysRevC.97.025204} {\bibfield  {journal} {\bibinfo
  {journal} {Phys. Rev. C}\ }\textbf {\bibinfo {volume} {97}},\ \bibinfo
  {pages} {025204} (\bibinfo {year} {2018})},\ \Eprint
  {https://arxiv.org/abs/1801.01536} {arXiv:1801.01536 [nucl-ex]} \BibitemShut
  {NoStop}%
\bibitem [{\citenamefont {Dudek}\ \emph {et~al.}(2012)\citenamefont {Dudek}
  \emph {et~al.}}]{Dudek:2012vr}%
  \BibitemOpen
  \bibfield  {author} {\bibinfo {author} {\bibfnamefont {J.}~\bibnamefont
  {Dudek}} \emph {et~al.},\ }\href {https://doi.org/10.1140/epja/i2012-12187-1}
  {\bibfield  {journal} {\bibinfo  {journal} {Eur. Phys. J. A}\ }\textbf
  {\bibinfo {volume} {48}},\ \bibinfo {pages} {187} (\bibinfo {year} {2012})},\
  \Eprint {https://arxiv.org/abs/1208.1244} {arXiv:1208.1244 [hep-ex]}
  \BibitemShut {NoStop}%
\bibitem [{\citenamefont {Arrington}\ \emph {et~al.}(2022)\citenamefont
  {Arrington} \emph {et~al.}}]{Arrington:2021alx}%
  \BibitemOpen
  \bibfield  {author} {\bibinfo {author} {\bibfnamefont {J.}~\bibnamefont
  {Arrington}} \emph {et~al.},\ }\href
  {https://doi.org/10.1016/j.ppnp.2022.103985} {\bibfield  {journal} {\bibinfo
  {journal} {Prog. Part. Nucl. Phys.}\ }\textbf {\bibinfo {volume} {127}},\
  \bibinfo {pages} {103985} (\bibinfo {year} {2022})},\ \Eprint
  {https://arxiv.org/abs/2112.00060} {arXiv:2112.00060 [nucl-ex]} \BibitemShut
  {NoStop}%
\bibitem [{\citenamefont {Arrington}\ \emph {et~al.}(2021)\citenamefont
  {Arrington} \emph {et~al.}}]{Arrington:2021biu}%
  \BibitemOpen
  \bibfield  {author} {\bibinfo {author} {\bibfnamefont {J.}~\bibnamefont
  {Arrington}} \emph {et~al.},\ }\href
  {https://doi.org/10.1088/1361-6471/abf5c3} {\bibfield  {journal} {\bibinfo
  {journal} {J. Phys. G}\ }\textbf {\bibinfo {volume} {48}},\ \bibinfo {pages}
  {075106} (\bibinfo {year} {2021})},\ \Eprint
  {https://arxiv.org/abs/2102.11788} {arXiv:2102.11788 [nucl-ex]} \BibitemShut
  {NoStop}%
\bibitem [{\citenamefont {Anderle}\ \emph {et~al.}(2021)\citenamefont {Anderle}
  \emph {et~al.}}]{Anderle:2021wcy}%
  \BibitemOpen
  \bibfield  {author} {\bibinfo {author} {\bibfnamefont {D.~P.}\ \bibnamefont
  {Anderle}} \emph {et~al.},\ }\href
  {https://doi.org/10.1007/s11467-021-1062-0} {\bibfield  {journal} {\bibinfo
  {journal} {Front. Phys. (Beijing)}\ }\textbf {\bibinfo {volume} {16}},\
  \bibinfo {pages} {64701} (\bibinfo {year} {2021})},\ \Eprint
  {https://arxiv.org/abs/2102.09222} {arXiv:2102.09222 [nucl-ex]} \BibitemShut
  {NoStop}%
\bibitem [{\citenamefont {Br\"ommel}\ \emph {et~al.}(2007)\citenamefont
  {Br\"ommel} \emph {et~al.}}]{QCDSFUKQCD:2006gmg}%
  \BibitemOpen
  \bibfield  {author} {\bibinfo {author} {\bibfnamefont {D.}~\bibnamefont
  {Br\"ommel}} \emph {et~al.} (\bibinfo {collaboration} {QCDSF/UKQCD}),\ }\href
  {https://doi.org/10.1140/epjc/s10052-007-0295-6} {\bibfield  {journal}
  {\bibinfo  {journal} {Eur. Phys. J. C}\ }\textbf {\bibinfo {volume} {51}},\
  \bibinfo {pages} {335} (\bibinfo {year} {2007})},\ \Eprint
  {https://arxiv.org/abs/hep-lat/0608021} {arXiv:hep-lat/0608021} \BibitemShut
  {NoStop}%
\bibitem [{\citenamefont {Boyle}\ \emph {et~al.}(2008)\citenamefont {Boyle},
  \citenamefont {Flynn}, \citenamefont {Juttner}, \citenamefont {Kelly},
  \citenamefont {de~Lima}, \citenamefont {Maynard}, \citenamefont {Sachrajda},\
  and\ \citenamefont {Zanotti}}]{Boyle:2008yd}%
  \BibitemOpen
  \bibfield  {author} {\bibinfo {author} {\bibfnamefont {P.~A.}\ \bibnamefont
  {Boyle}}, \bibinfo {author} {\bibfnamefont {J.~M.}\ \bibnamefont {Flynn}},
  \bibinfo {author} {\bibfnamefont {A.}~\bibnamefont {Juttner}}, \bibinfo
  {author} {\bibfnamefont {C.}~\bibnamefont {Kelly}}, \bibinfo {author}
  {\bibfnamefont {H.~P.}\ \bibnamefont {de~Lima}}, \bibinfo {author}
  {\bibfnamefont {C.~M.}\ \bibnamefont {Maynard}}, \bibinfo {author}
  {\bibfnamefont {C.~T.}\ \bibnamefont {Sachrajda}},\ and\ \bibinfo {author}
  {\bibfnamefont {J.~M.}\ \bibnamefont {Zanotti}},\ }\href
  {https://doi.org/10.1088/1126-6708/2008/07/112} {\bibfield  {journal}
  {\bibinfo  {journal} {JHEP}\ }\textbf {\bibinfo {volume} {07}},\ \bibinfo
  {pages} {112}},\ \Eprint {https://arxiv.org/abs/0804.3971} {arXiv:0804.3971
  [hep-lat]} \BibitemShut {NoStop}%
\bibitem [{\citenamefont {Aoki}\ \emph {et~al.}(2009)\citenamefont {Aoki} \emph
  {et~al.}}]{JLQCD:2009ofg}%
  \BibitemOpen
  \bibfield  {author} {\bibinfo {author} {\bibfnamefont {S.}~\bibnamefont
  {Aoki}} \emph {et~al.} (\bibinfo {collaboration} {JLQCD, TWQCD}),\ }\href
  {https://doi.org/10.1103/PhysRevD.80.034508} {\bibfield  {journal} {\bibinfo
  {journal} {Phys. Rev. D}\ }\textbf {\bibinfo {volume} {80}},\ \bibinfo
  {pages} {034508} (\bibinfo {year} {2009})},\ \Eprint
  {https://arxiv.org/abs/0905.2465} {arXiv:0905.2465 [hep-lat]} \BibitemShut
  {NoStop}%
\bibitem [{\citenamefont {Bali}\ \emph {et~al.}(2014)\citenamefont {Bali},
  \citenamefont {Collins}, \citenamefont {Gl\"assle}, \citenamefont
  {G\"ockeler}, \citenamefont {Javadi-Motaghi}, \citenamefont {Najjar},
  \citenamefont {S\"oldner},\ and\ \citenamefont {Sternbeck}}]{Bali:2013gya}%
  \BibitemOpen
  \bibfield  {author} {\bibinfo {author} {\bibfnamefont {G.}~\bibnamefont
  {Bali}}, \bibinfo {author} {\bibfnamefont {S.}~\bibnamefont {Collins}},
  \bibinfo {author} {\bibfnamefont {B.}~\bibnamefont {Gl\"assle}}, \bibinfo
  {author} {\bibfnamefont {M.}~\bibnamefont {G\"ockeler}}, \bibinfo {author}
  {\bibfnamefont {N.}~\bibnamefont {Javadi-Motaghi}}, \bibinfo {author}
  {\bibfnamefont {J.}~\bibnamefont {Najjar}}, \bibinfo {author} {\bibfnamefont
  {W.}~\bibnamefont {S\"oldner}},\ and\ \bibinfo {author} {\bibfnamefont
  {A.}~\bibnamefont {Sternbeck}},\ }\href {https://doi.org/10.22323/1.187.0447}
  {\bibfield  {journal} {\bibinfo  {journal} {PoS}\ }\textbf {\bibinfo {volume}
  {LATTICE2013}},\ \bibinfo {pages} {447} (\bibinfo {year} {2014})},\ \Eprint
  {https://arxiv.org/abs/1311.7639} {arXiv:1311.7639 [hep-lat]} \BibitemShut
  {NoStop}%
\bibitem [{\citenamefont {Fukaya}\ \emph {et~al.}(2014)\citenamefont {Fukaya},
  \citenamefont {Aoki}, \citenamefont {Hashimoto}, \citenamefont {Kaneko},
  \citenamefont {Matsufuru},\ and\ \citenamefont {Noaki}}]{Fukaya:2014jka}%
  \BibitemOpen
  \bibfield  {author} {\bibinfo {author} {\bibfnamefont {H.}~\bibnamefont
  {Fukaya}}, \bibinfo {author} {\bibfnamefont {S.}~\bibnamefont {Aoki}},
  \bibinfo {author} {\bibfnamefont {S.}~\bibnamefont {Hashimoto}}, \bibinfo
  {author} {\bibfnamefont {T.}~\bibnamefont {Kaneko}}, \bibinfo {author}
  {\bibfnamefont {H.}~\bibnamefont {Matsufuru}},\ and\ \bibinfo {author}
  {\bibfnamefont {J.}~\bibnamefont {Noaki}},\ }\href
  {https://doi.org/10.1103/PhysRevD.90.034506} {\bibfield  {journal} {\bibinfo
  {journal} {Phys. Rev. D}\ }\textbf {\bibinfo {volume} {90}},\ \bibinfo
  {pages} {034506} (\bibinfo {year} {2014})},\ \Eprint
  {https://arxiv.org/abs/1405.4077} {arXiv:1405.4077 [hep-lat]} \BibitemShut
  {NoStop}%
\bibitem [{\citenamefont {Colangelo}\ \emph {et~al.}(2019)\citenamefont
  {Colangelo}, \citenamefont {Hoferichter},\ and\ \citenamefont
  {Stoffer}}]{Colangelo:2018mtw}%
  \BibitemOpen
  \bibfield  {author} {\bibinfo {author} {\bibfnamefont {G.}~\bibnamefont
  {Colangelo}}, \bibinfo {author} {\bibfnamefont {M.}~\bibnamefont
  {Hoferichter}},\ and\ \bibinfo {author} {\bibfnamefont {P.}~\bibnamefont
  {Stoffer}},\ }\href {https://doi.org/10.1007/JHEP02(2019)006} {\bibfield
  {journal} {\bibinfo  {journal} {JHEP}\ }\textbf {\bibinfo {volume} {02}},\
  \bibinfo {pages} {006}},\ \Eprint {https://arxiv.org/abs/1810.00007}
  {arXiv:1810.00007 [hep-ph]} \BibitemShut {NoStop}%
\bibitem [{\citenamefont {Wang}\ \emph {et~al.}(2021)\citenamefont {Wang},
  \citenamefont {Liang}, \citenamefont {Draper}, \citenamefont {Liu},\ and\
  \citenamefont {Yang}}]{Wang:2020nbf}%
  \BibitemOpen
  \bibfield  {author} {\bibinfo {author} {\bibfnamefont {G.}~\bibnamefont
  {Wang}}, \bibinfo {author} {\bibfnamefont {J.}~\bibnamefont {Liang}},
  \bibinfo {author} {\bibfnamefont {T.}~\bibnamefont {Draper}}, \bibinfo
  {author} {\bibfnamefont {K.-F.}\ \bibnamefont {Liu}},\ and\ \bibinfo {author}
  {\bibfnamefont {Y.-B.}\ \bibnamefont {Yang}} (\bibinfo {collaboration}
  {chiQCD}),\ }\href {https://doi.org/10.1103/PhysRevD.104.074502} {\bibfield
  {journal} {\bibinfo  {journal} {Phys. Rev. D}\ }\textbf {\bibinfo {volume}
  {104}},\ \bibinfo {pages} {074502} (\bibinfo {year} {2021})},\ \Eprint
  {https://arxiv.org/abs/2006.05431} {arXiv:2006.05431 [hep-ph]} \BibitemShut
  {NoStop}%
\bibitem [{\citenamefont {Gao}\ \emph {et~al.}(2021)\citenamefont {Gao},
  \citenamefont {Karthik}, \citenamefont {Mukherjee}, \citenamefont
  {Petreczky}, \citenamefont {Syritsyn},\ and\ \citenamefont
  {Zhao}}]{Gao:2021xsm}%
  \BibitemOpen
  \bibfield  {author} {\bibinfo {author} {\bibfnamefont {X.}~\bibnamefont
  {Gao}}, \bibinfo {author} {\bibfnamefont {N.}~\bibnamefont {Karthik}},
  \bibinfo {author} {\bibfnamefont {S.}~\bibnamefont {Mukherjee}}, \bibinfo
  {author} {\bibfnamefont {P.}~\bibnamefont {Petreczky}}, \bibinfo {author}
  {\bibfnamefont {S.}~\bibnamefont {Syritsyn}},\ and\ \bibinfo {author}
  {\bibfnamefont {Y.}~\bibnamefont {Zhao}},\ }\href
  {https://doi.org/10.1103/PhysRevD.104.114515} {\bibfield  {journal} {\bibinfo
   {journal} {Phys. Rev. D}\ }\textbf {\bibinfo {volume} {104}},\ \bibinfo
  {pages} {114515} (\bibinfo {year} {2021})},\ \Eprint
  {https://arxiv.org/abs/2102.06047} {arXiv:2102.06047 [hep-lat]} \BibitemShut
  {NoStop}%
\bibitem [{\citenamefont {Koponen}\ \emph {et~al.}(2016)\citenamefont
  {Koponen}, \citenamefont {Bursa}, \citenamefont {Davies}, \citenamefont
  {Dowdall},\ and\ \citenamefont {Lepage}}]{Koponen:2015tkr}%
  \BibitemOpen
  \bibfield  {author} {\bibinfo {author} {\bibfnamefont {J.}~\bibnamefont
  {Koponen}}, \bibinfo {author} {\bibfnamefont {F.}~\bibnamefont {Bursa}},
  \bibinfo {author} {\bibfnamefont {C.~T.~H.}\ \bibnamefont {Davies}}, \bibinfo
  {author} {\bibfnamefont {R.~J.}\ \bibnamefont {Dowdall}},\ and\ \bibinfo
  {author} {\bibfnamefont {G.~P.}\ \bibnamefont {Lepage}},\ }\href
  {https://doi.org/10.1103/PhysRevD.93.054503} {\bibfield  {journal} {\bibinfo
  {journal} {Phys. Rev. D}\ }\textbf {\bibinfo {volume} {93}},\ \bibinfo
  {pages} {054503} (\bibinfo {year} {2016})},\ \Eprint
  {https://arxiv.org/abs/1511.07382} {arXiv:1511.07382 [hep-lat]} \BibitemShut
  {NoStop}%
\bibitem [{\citenamefont {Alexandrou}\ \emph {et~al.}(2018)\citenamefont
  {Alexandrou} \emph {et~al.}}]{ETM:2017wqc}%
  \BibitemOpen
  \bibfield  {author} {\bibinfo {author} {\bibfnamefont {C.}~\bibnamefont
  {Alexandrou}} \emph {et~al.} (\bibinfo {collaboration} {ETM}),\ }\href
  {https://doi.org/10.1103/PhysRevD.97.014508} {\bibfield  {journal} {\bibinfo
  {journal} {Phys. Rev. D}\ }\textbf {\bibinfo {volume} {97}},\ \bibinfo
  {pages} {014508} (\bibinfo {year} {2018})},\ \Eprint
  {https://arxiv.org/abs/1710.10401} {arXiv:1710.10401 [hep-lat]} \BibitemShut
  {NoStop}%
\bibitem [{\citenamefont {Kaneko}\ \emph {et~al.}(2010)\citenamefont {Kaneko},
  \citenamefont {Aoki}, \citenamefont {Cossu}, \citenamefont {Fukaya},
  \citenamefont {Hashimoto}, \citenamefont {Noaki},\ and\ \citenamefont
  {Onogi}}]{Kaneko:2010ru}%
  \BibitemOpen
  \bibfield  {author} {\bibinfo {author} {\bibfnamefont {T.}~\bibnamefont
  {Kaneko}}, \bibinfo {author} {\bibfnamefont {S.}~\bibnamefont {Aoki}},
  \bibinfo {author} {\bibfnamefont {G.}~\bibnamefont {Cossu}}, \bibinfo
  {author} {\bibfnamefont {H.}~\bibnamefont {Fukaya}}, \bibinfo {author}
  {\bibfnamefont {S.}~\bibnamefont {Hashimoto}}, \bibinfo {author}
  {\bibfnamefont {J.}~\bibnamefont {Noaki}},\ and\ \bibinfo {author}
  {\bibfnamefont {T.}~\bibnamefont {Onogi}} (\bibinfo {collaboration}
  {JLQCD}),\ }\href {https://doi.org/10.22323/1.105.0146} {\bibfield  {journal}
  {\bibinfo  {journal} {PoS}\ }\textbf {\bibinfo {volume} {LATTICE2010}},\
  \bibinfo {pages} {146} (\bibinfo {year} {2010})},\ \Eprint
  {https://arxiv.org/abs/1012.0137} {arXiv:1012.0137 [hep-lat]} \BibitemShut
  {NoStop}%
\bibitem [{\citenamefont {Alexandrou}\ \emph {et~al.}(2022)\citenamefont
  {Alexandrou}, \citenamefont {Bacchio}, \citenamefont {Cloet}, \citenamefont
  {Constantinou}, \citenamefont {Delmar}, \citenamefont {Hadjiyiannakou},
  \citenamefont {Koutsou}, \citenamefont {Lauer},\ and\ \citenamefont
  {Vaquero}}]{Alexandrou:2021ztx}%
  \BibitemOpen
  \bibfield  {author} {\bibinfo {author} {\bibfnamefont {C.}~\bibnamefont
  {Alexandrou}}, \bibinfo {author} {\bibfnamefont {S.}~\bibnamefont {Bacchio}},
  \bibinfo {author} {\bibfnamefont {I.}~\bibnamefont {Cloet}}, \bibinfo
  {author} {\bibfnamefont {M.}~\bibnamefont {Constantinou}}, \bibinfo {author}
  {\bibfnamefont {J.}~\bibnamefont {Delmar}}, \bibinfo {author} {\bibfnamefont
  {K.}~\bibnamefont {Hadjiyiannakou}}, \bibinfo {author} {\bibfnamefont
  {G.}~\bibnamefont {Koutsou}}, \bibinfo {author} {\bibfnamefont
  {C.}~\bibnamefont {Lauer}},\ and\ \bibinfo {author} {\bibfnamefont
  {A.}~\bibnamefont {Vaquero}} (\bibinfo {collaboration} {ETM}),\ }\href
  {https://doi.org/10.1103/PhysRevD.105.054502} {\bibfield  {journal} {\bibinfo
   {journal} {Phys. Rev. D}\ }\textbf {\bibinfo {volume} {105}},\ \bibinfo
  {pages} {054502} (\bibinfo {year} {2022})},\ \Eprint
  {https://arxiv.org/abs/2111.08135} {arXiv:2111.08135 [hep-lat]} \BibitemShut
  {NoStop}%
\bibitem [{\citenamefont {Chambers}\ \emph {et~al.}(2017)\citenamefont
  {Chambers} \emph {et~al.}}]{QCDSF:2017ssq}%
  \BibitemOpen
  \bibfield  {author} {\bibinfo {author} {\bibfnamefont {A.~J.}\ \bibnamefont
  {Chambers}} \emph {et~al.} (\bibinfo {collaboration} {QCDSF, UKQCD, CSSM}),\
  }\href {https://doi.org/10.1103/PhysRevD.96.114509} {\bibfield  {journal}
  {\bibinfo  {journal} {Phys. Rev. D}\ }\textbf {\bibinfo {volume} {96}},\
  \bibinfo {pages} {114509} (\bibinfo {year} {2017})},\ \Eprint
  {https://arxiv.org/abs/1702.01513} {arXiv:1702.01513 [hep-lat]} \BibitemShut
  {NoStop}%
\bibitem [{\citenamefont {Koponen}\ \emph {et~al.}(2017)\citenamefont
  {Koponen}, \citenamefont {Zimermmane-Santos}, \citenamefont {Davies},
  \citenamefont {Lepage},\ and\ \citenamefont {Lytle}}]{Koponen:2017fvm}%
  \BibitemOpen
  \bibfield  {author} {\bibinfo {author} {\bibfnamefont {J.}~\bibnamefont
  {Koponen}}, \bibinfo {author} {\bibfnamefont {A.~C.}\ \bibnamefont
  {Zimermmane-Santos}}, \bibinfo {author} {\bibfnamefont {C.~T.~H.}\
  \bibnamefont {Davies}}, \bibinfo {author} {\bibfnamefont {G.~P.}\
  \bibnamefont {Lepage}},\ and\ \bibinfo {author} {\bibfnamefont {A.~T.}\
  \bibnamefont {Lytle}},\ }\href {https://doi.org/10.1103/PhysRevD.96.054501}
  {\bibfield  {journal} {\bibinfo  {journal} {Phys. Rev. D}\ }\textbf {\bibinfo
  {volume} {96}},\ \bibinfo {pages} {054501} (\bibinfo {year} {2017})},\
  \Eprint {https://arxiv.org/abs/1701.04250} {arXiv:1701.04250 [hep-lat]}
  \BibitemShut {NoStop}%
\bibitem [{\citenamefont {Davies}\ \emph {et~al.}(2018)\citenamefont {Davies},
  \citenamefont {Koponen}, \citenamefont {Lepage}, \citenamefont {Lytle},\ and\
  \citenamefont {Zimermmane-Santos}}]{Davies:2018zav}%
  \BibitemOpen
  \bibfield  {author} {\bibinfo {author} {\bibfnamefont {C.~T.~H.}\
  \bibnamefont {Davies}}, \bibinfo {author} {\bibfnamefont {J.}~\bibnamefont
  {Koponen}}, \bibinfo {author} {\bibfnamefont {P.~G.}\ \bibnamefont {Lepage}},
  \bibinfo {author} {\bibfnamefont {A.~T.}\ \bibnamefont {Lytle}},\ and\
  \bibinfo {author} {\bibfnamefont {A.~C.}\ \bibnamefont {Zimermmane-Santos}}
  (\bibinfo {collaboration} {HPQCD}),\ }\href
  {https://doi.org/10.22323/1.334.0298} {\bibfield  {journal} {\bibinfo
  {journal} {PoS}\ }\textbf {\bibinfo {volume} {LATTICE2018}},\ \bibinfo
  {pages} {298} (\bibinfo {year} {2018})},\ \Eprint
  {https://arxiv.org/abs/1902.03808} {arXiv:1902.03808 [hep-lat]} \BibitemShut
  {NoStop}%
\bibitem [{\citenamefont {Bali}\ \emph {et~al.}(2019)\citenamefont {Bali},
  \citenamefont {Braun}, \citenamefont {B\"urger}, \citenamefont {G\"ockeler},
  \citenamefont {Gruber}, \citenamefont {Hutzler}, \citenamefont {Korcyl},
  \citenamefont {Sch\"afer}, \citenamefont {Sternbeck},\ and\ \citenamefont
  {Wein}}]{RQCD:2019osh}%
  \BibitemOpen
  \bibfield  {author} {\bibinfo {author} {\bibfnamefont {G.~S.}\ \bibnamefont
  {Bali}}, \bibinfo {author} {\bibfnamefont {V.~M.}\ \bibnamefont {Braun}},
  \bibinfo {author} {\bibfnamefont {S.}~\bibnamefont {B\"urger}}, \bibinfo
  {author} {\bibfnamefont {M.}~\bibnamefont {G\"ockeler}}, \bibinfo {author}
  {\bibfnamefont {M.}~\bibnamefont {Gruber}}, \bibinfo {author} {\bibfnamefont
  {F.}~\bibnamefont {Hutzler}}, \bibinfo {author} {\bibfnamefont
  {P.}~\bibnamefont {Korcyl}}, \bibinfo {author} {\bibfnamefont
  {A.}~\bibnamefont {Sch\"afer}}, \bibinfo {author} {\bibfnamefont
  {A.}~\bibnamefont {Sternbeck}},\ and\ \bibinfo {author} {\bibfnamefont
  {P.}~\bibnamefont {Wein}} (\bibinfo {collaboration} {RQCD}),\ }\href
  {https://doi.org/10.1007/JHEP08(2019)065} {\bibfield  {journal} {\bibinfo
  {journal} {JHEP}\ }\textbf {\bibinfo {volume} {08}},\ \bibinfo {pages}
  {065}},\ \bibinfo {note} {[Addendum: JHEP 11, 037 (2020)]},\ \Eprint
  {https://arxiv.org/abs/1903.08038} {arXiv:1903.08038 [hep-lat]} \BibitemShut
  {NoStop}%
\bibitem [{\citenamefont {Detmold}\ \emph {et~al.}(2022)\citenamefont
  {Detmold}, \citenamefont {Grebe}, \citenamefont {Kanamori}, \citenamefont
  {Lin}, \citenamefont {Mondal}, \citenamefont {Perry},\ and\ \citenamefont
  {Zhao}}]{Detmold:2021qln}%
  \BibitemOpen
  \bibfield  {author} {\bibinfo {author} {\bibfnamefont {W.}~\bibnamefont
  {Detmold}}, \bibinfo {author} {\bibfnamefont {A.~V.}\ \bibnamefont {Grebe}},
  \bibinfo {author} {\bibfnamefont {I.}~\bibnamefont {Kanamori}}, \bibinfo
  {author} {\bibfnamefont {C.~J.~D.}\ \bibnamefont {Lin}}, \bibinfo {author}
  {\bibfnamefont {S.}~\bibnamefont {Mondal}}, \bibinfo {author} {\bibfnamefont
  {R.~J.}\ \bibnamefont {Perry}},\ and\ \bibinfo {author} {\bibfnamefont
  {Y.}~\bibnamefont {Zhao}} (\bibinfo {collaboration} {HOPE}),\ }\href
  {https://doi.org/10.1103/PhysRevD.105.034506} {\bibfield  {journal} {\bibinfo
   {journal} {Phys. Rev. D}\ }\textbf {\bibinfo {volume} {105}},\ \bibinfo
  {pages} {034506} (\bibinfo {year} {2022})},\ \Eprint
  {https://arxiv.org/abs/2109.15241} {arXiv:2109.15241 [hep-lat]} \BibitemShut
  {NoStop}%
\bibitem [{\citenamefont {Hua}\ \emph {et~al.}(2022)\citenamefont {Hua} \emph
  {et~al.}}]{LatticeParton:2022zqc}%
  \BibitemOpen
  \bibfield  {author} {\bibinfo {author} {\bibfnamefont {J.}~\bibnamefont
  {Hua}} \emph {et~al.} (\bibinfo {collaboration} {Lattice Parton}),\ }\href
  {https://doi.org/10.1103/PhysRevLett.129.132001} {\bibfield  {journal}
  {\bibinfo  {journal} {Phys. Rev. Lett.}\ }\textbf {\bibinfo {volume} {129}},\
  \bibinfo {pages} {132001} (\bibinfo {year} {2022})},\ \Eprint
  {https://arxiv.org/abs/2201.09173} {arXiv:2201.09173 [hep-lat]} \BibitemShut
  {NoStop}%
\bibitem [{\citenamefont {Gao}\ \emph {et~al.}(2022{\natexlab{b}})\citenamefont
  {Gao}, \citenamefont {Hanlon}, \citenamefont {Karthik}, \citenamefont
  {Mukherjee}, \citenamefont {Petreczky}, \citenamefont {Scior}, \citenamefont
  {Syritsyn},\ and\ \citenamefont {Zhao}}]{Gao:2022vyh}%
  \BibitemOpen
  \bibfield  {author} {\bibinfo {author} {\bibfnamefont {X.}~\bibnamefont
  {Gao}}, \bibinfo {author} {\bibfnamefont {A.~D.}\ \bibnamefont {Hanlon}},
  \bibinfo {author} {\bibfnamefont {N.}~\bibnamefont {Karthik}}, \bibinfo
  {author} {\bibfnamefont {S.}~\bibnamefont {Mukherjee}}, \bibinfo {author}
  {\bibfnamefont {P.}~\bibnamefont {Petreczky}}, \bibinfo {author}
  {\bibfnamefont {P.}~\bibnamefont {Scior}}, \bibinfo {author} {\bibfnamefont
  {S.}~\bibnamefont {Syritsyn}},\ and\ \bibinfo {author} {\bibfnamefont
  {Y.}~\bibnamefont {Zhao}},\ }\href
  {https://doi.org/10.1103/PhysRevD.106.074505} {\bibfield  {journal} {\bibinfo
   {journal} {Phys. Rev. D}\ }\textbf {\bibinfo {volume} {106}},\ \bibinfo
  {pages} {074505} (\bibinfo {year} {2022}{\natexlab{b}})},\ \Eprint
  {https://arxiv.org/abs/2206.04084} {arXiv:2206.04084 [hep-lat]} \BibitemShut
  {NoStop}%
\bibitem [{\citenamefont {Baker}\ \emph {et~al.}(2024)\citenamefont {Baker},
  \citenamefont {Bollweg}, \citenamefont {Boyle}, \citenamefont {Clo\"et},
  \citenamefont {Gao}, \citenamefont {Mukherjee}, \citenamefont {Petreczky},
  \citenamefont {Zhang},\ and\ \citenamefont {Zhao}}]{Baker:2024zcd}%
  \BibitemOpen
  \bibfield  {author} {\bibinfo {author} {\bibfnamefont {E.}~\bibnamefont
  {Baker}}, \bibinfo {author} {\bibfnamefont {D.}~\bibnamefont {Bollweg}},
  \bibinfo {author} {\bibfnamefont {P.}~\bibnamefont {Boyle}}, \bibinfo
  {author} {\bibfnamefont {I.}~\bibnamefont {Clo\"et}}, \bibinfo {author}
  {\bibfnamefont {X.}~\bibnamefont {Gao}}, \bibinfo {author} {\bibfnamefont
  {S.}~\bibnamefont {Mukherjee}}, \bibinfo {author} {\bibfnamefont
  {P.}~\bibnamefont {Petreczky}}, \bibinfo {author} {\bibfnamefont
  {R.}~\bibnamefont {Zhang}},\ and\ \bibinfo {author} {\bibfnamefont
  {Y.}~\bibnamefont {Zhao}},\ }\href@noop {} {\  (\bibinfo {year} {2024})},\
  \Eprint {https://arxiv.org/abs/2405.20120} {arXiv:2405.20120 [hep-lat]}
  \BibitemShut {NoStop}%
\bibitem [{\citenamefont {Cloet}\ \emph {et~al.}(2024)\citenamefont {Cloet},
  \citenamefont {Gao}, \citenamefont {Mukherjee}, \citenamefont {Syritsyn},
  \citenamefont {Karthik}, \citenamefont {Petreczky}, \citenamefont {Zhang},\
  and\ \citenamefont {Zhao}}]{Cloet:2024vbv}%
  \BibitemOpen
  \bibfield  {author} {\bibinfo {author} {\bibfnamefont {I.}~\bibnamefont
  {Cloet}}, \bibinfo {author} {\bibfnamefont {X.}~\bibnamefont {Gao}}, \bibinfo
  {author} {\bibfnamefont {S.}~\bibnamefont {Mukherjee}}, \bibinfo {author}
  {\bibfnamefont {S.}~\bibnamefont {Syritsyn}}, \bibinfo {author}
  {\bibfnamefont {N.}~\bibnamefont {Karthik}}, \bibinfo {author} {\bibfnamefont
  {P.}~\bibnamefont {Petreczky}}, \bibinfo {author} {\bibfnamefont
  {R.}~\bibnamefont {Zhang}},\ and\ \bibinfo {author} {\bibfnamefont
  {Y.}~\bibnamefont {Zhao}},\ }\href@noop {} {\  (\bibinfo {year} {2024})},\
  \Eprint {https://arxiv.org/abs/2407.00206} {arXiv:2407.00206 [hep-lat]}
  \BibitemShut {NoStop}%
\bibitem [{\citenamefont {Chen}\ \emph {et~al.}(2023)\citenamefont {Chen},
  \citenamefont {Chen}, \citenamefont {Feng},\ and\ \citenamefont
  {Jia}}]{Chen:2023byr}%
  \BibitemOpen
  \bibfield  {author} {\bibinfo {author} {\bibfnamefont {L.-B.}\ \bibnamefont
  {Chen}}, \bibinfo {author} {\bibfnamefont {W.}~\bibnamefont {Chen}}, \bibinfo
  {author} {\bibfnamefont {F.}~\bibnamefont {Feng}},\ and\ \bibinfo {author}
  {\bibfnamefont {Y.}~\bibnamefont {Jia}},\ }\href@noop {} {\  (\bibinfo {year}
  {2023})},\ \Eprint {https://arxiv.org/abs/2312.17228} {arXiv:2312.17228
  [hep-ph]} \BibitemShut {NoStop}%
\bibitem [{\citenamefont {Bazavov}\ \emph {et~al.}(2019)\citenamefont {Bazavov}
  \emph {et~al.}}]{Bazavov:2019www}%
  \BibitemOpen
  \bibfield  {author} {\bibinfo {author} {\bibfnamefont {A.}~\bibnamefont
  {Bazavov}} \emph {et~al.},\ }\href
  {https://doi.org/10.1103/PhysRevD.100.094510} {\bibfield  {journal} {\bibinfo
   {journal} {Phys. Rev. D}\ }\textbf {\bibinfo {volume} {100}},\ \bibinfo
  {pages} {094510} (\bibinfo {year} {2019})},\ \Eprint
  {https://arxiv.org/abs/1908.09552} {arXiv:1908.09552 [hep-lat]} \BibitemShut
  {NoStop}%
\bibitem [{\citenamefont {Follana}\ \emph {et~al.}(2007)\citenamefont
  {Follana}, \citenamefont {Mason}, \citenamefont {Davies}, \citenamefont
  {Hornbostel}, \citenamefont {Lepage}, \citenamefont {Shigemitsu},
  \citenamefont {Trottier},\ and\ \citenamefont {Wong}}]{Follana:2006rc}%
  \BibitemOpen
  \bibfield  {author} {\bibinfo {author} {\bibfnamefont {E.}~\bibnamefont
  {Follana}}, \bibinfo {author} {\bibfnamefont {Q.}~\bibnamefont {Mason}},
  \bibinfo {author} {\bibfnamefont {C.}~\bibnamefont {Davies}}, \bibinfo
  {author} {\bibfnamefont {K.}~\bibnamefont {Hornbostel}}, \bibinfo {author}
  {\bibfnamefont {G.~P.}\ \bibnamefont {Lepage}}, \bibinfo {author}
  {\bibfnamefont {J.}~\bibnamefont {Shigemitsu}}, \bibinfo {author}
  {\bibfnamefont {H.}~\bibnamefont {Trottier}},\ and\ \bibinfo {author}
  {\bibfnamefont {K.}~\bibnamefont {Wong}} (\bibinfo {collaboration} {HPQCD,
  UKQCD}),\ }\href {https://doi.org/10.1103/PhysRevD.75.054502} {\bibfield
  {journal} {\bibinfo  {journal} {Phys. Rev. D}\ }\textbf {\bibinfo {volume}
  {75}},\ \bibinfo {pages} {054502} (\bibinfo {year} {2007})},\ \Eprint
  {https://arxiv.org/abs/hep-lat/0610092} {arXiv:hep-lat/0610092} \BibitemShut
  {NoStop}%
\bibitem [{\citenamefont {Hasenfratz}\ and\ \citenamefont
  {Knechtli}(2001)}]{Hasenfratz:2001hp}%
  \BibitemOpen
  \bibfield  {author} {\bibinfo {author} {\bibfnamefont {A.}~\bibnamefont
  {Hasenfratz}}\ and\ \bibinfo {author} {\bibfnamefont {F.}~\bibnamefont
  {Knechtli}},\ }\href {https://doi.org/10.1103/PhysRevD.64.034504} {\bibfield
  {journal} {\bibinfo  {journal} {Phys. Rev. D}\ }\textbf {\bibinfo {volume}
  {64}},\ \bibinfo {pages} {034504} (\bibinfo {year} {2001})},\ \Eprint
  {https://arxiv.org/abs/hep-lat/0103029} {arXiv:hep-lat/0103029} \BibitemShut
  {NoStop}%
\bibitem [{\citenamefont {Gao}\ \emph {et~al.}(2022{\natexlab{c}})\citenamefont
  {Gao}, \citenamefont {Hanlon}, \citenamefont {Karthik}, \citenamefont
  {Mukherjee}, \citenamefont {Petreczky}, \citenamefont {Scior}, \citenamefont
  {Shi}, \citenamefont {Syritsyn}, \citenamefont {Zhao},\ and\ \citenamefont
  {Zhou}}]{Gao:2022iex}%
  \BibitemOpen
  \bibfield  {author} {\bibinfo {author} {\bibfnamefont {X.}~\bibnamefont
  {Gao}}, \bibinfo {author} {\bibfnamefont {A.~D.}\ \bibnamefont {Hanlon}},
  \bibinfo {author} {\bibfnamefont {N.}~\bibnamefont {Karthik}}, \bibinfo
  {author} {\bibfnamefont {S.}~\bibnamefont {Mukherjee}}, \bibinfo {author}
  {\bibfnamefont {P.}~\bibnamefont {Petreczky}}, \bibinfo {author}
  {\bibfnamefont {P.}~\bibnamefont {Scior}}, \bibinfo {author} {\bibfnamefont
  {S.}~\bibnamefont {Shi}}, \bibinfo {author} {\bibfnamefont {S.}~\bibnamefont
  {Syritsyn}}, \bibinfo {author} {\bibfnamefont {Y.}~\bibnamefont {Zhao}},\
  and\ \bibinfo {author} {\bibfnamefont {K.}~\bibnamefont {Zhou}},\ }\href
  {https://doi.org/10.1103/PhysRevD.106.114510} {\bibfield  {journal} {\bibinfo
   {journal} {Phys. Rev. D}\ }\textbf {\bibinfo {volume} {106}},\ \bibinfo
  {pages} {114510} (\bibinfo {year} {2022}{\natexlab{c}})},\ \Eprint
  {https://arxiv.org/abs/2208.02297} {arXiv:2208.02297 [hep-lat]} \BibitemShut
  {NoStop}%
\bibitem [{\citenamefont {Gao}\ \emph {et~al.}(2020)\citenamefont {Gao},
  \citenamefont {Jin}, \citenamefont {Kallidonis}, \citenamefont {Karthik},
  \citenamefont {Mukherjee}, \citenamefont {Petreczky}, \citenamefont
  {Shugert}, \citenamefont {Syritsyn},\ and\ \citenamefont
  {Zhao}}]{Gao:2020ito}%
  \BibitemOpen
  \bibfield  {author} {\bibinfo {author} {\bibfnamefont {X.}~\bibnamefont
  {Gao}}, \bibinfo {author} {\bibfnamefont {L.}~\bibnamefont {Jin}}, \bibinfo
  {author} {\bibfnamefont {C.}~\bibnamefont {Kallidonis}}, \bibinfo {author}
  {\bibfnamefont {N.}~\bibnamefont {Karthik}}, \bibinfo {author} {\bibfnamefont
  {S.}~\bibnamefont {Mukherjee}}, \bibinfo {author} {\bibfnamefont
  {P.}~\bibnamefont {Petreczky}}, \bibinfo {author} {\bibfnamefont
  {C.}~\bibnamefont {Shugert}}, \bibinfo {author} {\bibfnamefont
  {S.}~\bibnamefont {Syritsyn}},\ and\ \bibinfo {author} {\bibfnamefont
  {Y.}~\bibnamefont {Zhao}},\ }\href
  {https://doi.org/10.1103/PhysRevD.102.094513} {\bibfield  {journal} {\bibinfo
   {journal} {Phys. Rev. D}\ }\textbf {\bibinfo {volume} {102}},\ \bibinfo
  {pages} {094513} (\bibinfo {year} {2020})},\ \Eprint
  {https://arxiv.org/abs/2007.06590} {arXiv:2007.06590 [hep-lat]} \BibitemShut
  {NoStop}%
\bibitem [{\citenamefont {Brannick}\ \emph {et~al.}(2008)\citenamefont
  {Brannick}, \citenamefont {Brower}, \citenamefont {Clark}, \citenamefont
  {Osborn},\ and\ \citenamefont {Rebbi}}]{Brannick:2007ue}%
  \BibitemOpen
  \bibfield  {author} {\bibinfo {author} {\bibfnamefont {J.}~\bibnamefont
  {Brannick}}, \bibinfo {author} {\bibfnamefont {R.~C.}\ \bibnamefont
  {Brower}}, \bibinfo {author} {\bibfnamefont {M.~A.}\ \bibnamefont {Clark}},
  \bibinfo {author} {\bibfnamefont {J.~C.}\ \bibnamefont {Osborn}},\ and\
  \bibinfo {author} {\bibfnamefont {C.}~\bibnamefont {Rebbi}},\ }\href
  {https://doi.org/10.1103/PhysRevLett.100.041601} {\bibfield  {journal}
  {\bibinfo  {journal} {Phys. Rev. Lett.}\ }\textbf {\bibinfo {volume} {100}},\
  \bibinfo {pages} {041601} (\bibinfo {year} {2008})},\ \Eprint
  {https://arxiv.org/abs/0707.4018} {arXiv:0707.4018 [hep-lat]} \BibitemShut
  {NoStop}%
\bibitem [{\citenamefont {Clark}\ \emph {et~al.}(2010)\citenamefont {Clark},
  \citenamefont {Babich}, \citenamefont {Barros}, \citenamefont {Brower},\ and\
  \citenamefont {Rebbi}}]{Clark:2009wm}%
  \BibitemOpen
  \bibfield  {author} {\bibinfo {author} {\bibfnamefont {M.~A.}\ \bibnamefont
  {Clark}}, \bibinfo {author} {\bibfnamefont {R.}~\bibnamefont {Babich}},
  \bibinfo {author} {\bibfnamefont {K.}~\bibnamefont {Barros}}, \bibinfo
  {author} {\bibfnamefont {R.~C.}\ \bibnamefont {Brower}},\ and\ \bibinfo
  {author} {\bibfnamefont {C.}~\bibnamefont {Rebbi}} (\bibinfo {collaboration}
  {QUDA}),\ }\href {https://doi.org/10.1016/j.cpc.2010.05.002} {\bibfield
  {journal} {\bibinfo  {journal} {Comput. Phys. Commun.}\ }\textbf {\bibinfo
  {volume} {181}},\ \bibinfo {pages} {1517} (\bibinfo {year} {2010})},\ \Eprint
  {https://arxiv.org/abs/0911.3191} {arXiv:0911.3191 [hep-lat]} \BibitemShut
  {NoStop}%
\bibitem [{\citenamefont {Babich}\ \emph {et~al.}(2011)\citenamefont {Babich},
  \citenamefont {Clark}, \citenamefont {Joo}, \citenamefont {Shi},
  \citenamefont {Brower},\ and\ \citenamefont {Gottlieb}}]{Babich:2011np}%
  \BibitemOpen
  \bibfield  {author} {\bibinfo {author} {\bibfnamefont {R.}~\bibnamefont
  {Babich}}, \bibinfo {author} {\bibfnamefont {M.~A.}\ \bibnamefont {Clark}},
  \bibinfo {author} {\bibfnamefont {B.}~\bibnamefont {Joo}}, \bibinfo {author}
  {\bibfnamefont {G.}~\bibnamefont {Shi}}, \bibinfo {author} {\bibfnamefont
  {R.~C.}\ \bibnamefont {Brower}},\ and\ \bibinfo {author} {\bibfnamefont
  {S.}~\bibnamefont {Gottlieb}} (\bibinfo {collaboration} {QUDA})\ }(\bibinfo
  {year} {2011})\ \Eprint {https://arxiv.org/abs/1109.2935} {arXiv:1109.2935
  [hep-lat]} \BibitemShut {NoStop}%
\bibitem [{\citenamefont {Clark}\ \emph {et~al.}(2016)\citenamefont {Clark},
  \citenamefont {Jo\'o}, \citenamefont {Strelchenko}, \citenamefont {Cheng},
  \citenamefont {Gambhir},\ and\ \citenamefont {Brower}}]{Clark:2016rdz}%
  \BibitemOpen
  \bibfield  {author} {\bibinfo {author} {\bibfnamefont {M.~A.}\ \bibnamefont
  {Clark}}, \bibinfo {author} {\bibfnamefont {B.}~\bibnamefont {Jo\'o}},
  \bibinfo {author} {\bibfnamefont {A.}~\bibnamefont {Strelchenko}}, \bibinfo
  {author} {\bibfnamefont {M.}~\bibnamefont {Cheng}}, \bibinfo {author}
  {\bibfnamefont {A.}~\bibnamefont {Gambhir}},\ and\ \bibinfo {author}
  {\bibfnamefont {R.}~\bibnamefont {Brower}} (\bibinfo {collaboration}
  {QUDA}),\ }\href@noop {} {\  (\bibinfo {year} {2016})},\ \Eprint
  {https://arxiv.org/abs/1612.07873} {arXiv:1612.07873 [hep-lat]} \BibitemShut
  {NoStop}%
\bibitem [{\citenamefont {Shintani}\ \emph {et~al.}(2015)\citenamefont
  {Shintani}, \citenamefont {Arthur}, \citenamefont {Blum}, \citenamefont
  {Izubuchi}, \citenamefont {Jung},\ and\ \citenamefont
  {Lehner}}]{Shintani:2014vja}%
  \BibitemOpen
  \bibfield  {author} {\bibinfo {author} {\bibfnamefont {E.}~\bibnamefont
  {Shintani}}, \bibinfo {author} {\bibfnamefont {R.}~\bibnamefont {Arthur}},
  \bibinfo {author} {\bibfnamefont {T.}~\bibnamefont {Blum}}, \bibinfo {author}
  {\bibfnamefont {T.}~\bibnamefont {Izubuchi}}, \bibinfo {author}
  {\bibfnamefont {C.}~\bibnamefont {Jung}},\ and\ \bibinfo {author}
  {\bibfnamefont {C.}~\bibnamefont {Lehner}},\ }\href
  {https://doi.org/10.1103/PhysRevD.91.114511} {\bibfield  {journal} {\bibinfo
  {journal} {Phys. Rev. D}\ }\textbf {\bibinfo {volume} {91}},\ \bibinfo
  {pages} {114511} (\bibinfo {year} {2015})},\ \Eprint
  {https://arxiv.org/abs/1402.0244} {arXiv:1402.0244 [hep-lat]} \BibitemShut
  {NoStop}%
\bibitem [{\citenamefont {Izubuchi}\ \emph {et~al.}(2019)\citenamefont
  {Izubuchi}, \citenamefont {Jin}, \citenamefont {Kallidonis}, \citenamefont
  {Karthik}, \citenamefont {Mukherjee}, \citenamefont {Petreczky},
  \citenamefont {Shugert},\ and\ \citenamefont {Syritsyn}}]{Izubuchi:2019lyk}%
  \BibitemOpen
  \bibfield  {author} {\bibinfo {author} {\bibfnamefont {T.}~\bibnamefont
  {Izubuchi}}, \bibinfo {author} {\bibfnamefont {L.}~\bibnamefont {Jin}},
  \bibinfo {author} {\bibfnamefont {C.}~\bibnamefont {Kallidonis}}, \bibinfo
  {author} {\bibfnamefont {N.}~\bibnamefont {Karthik}}, \bibinfo {author}
  {\bibfnamefont {S.}~\bibnamefont {Mukherjee}}, \bibinfo {author}
  {\bibfnamefont {P.}~\bibnamefont {Petreczky}}, \bibinfo {author}
  {\bibfnamefont {C.}~\bibnamefont {Shugert}},\ and\ \bibinfo {author}
  {\bibfnamefont {S.}~\bibnamefont {Syritsyn}},\ }\href
  {https://doi.org/10.1103/PhysRevD.100.034516} {\bibfield  {journal} {\bibinfo
   {journal} {Phys. Rev. D}\ }\textbf {\bibinfo {volume} {100}},\ \bibinfo
  {pages} {034516} (\bibinfo {year} {2019})},\ \Eprint
  {https://arxiv.org/abs/1905.06349} {arXiv:1905.06349 [hep-lat]} \BibitemShut
  {NoStop}%
\bibitem [{\citenamefont {Bali}\ \emph {et~al.}(2016)\citenamefont {Bali},
  \citenamefont {Lang}, \citenamefont {Musch},\ and\ \citenamefont
  {Sch\"afer}}]{Bali:2016lva}%
  \BibitemOpen
  \bibfield  {author} {\bibinfo {author} {\bibfnamefont {G.~S.}\ \bibnamefont
  {Bali}}, \bibinfo {author} {\bibfnamefont {B.}~\bibnamefont {Lang}}, \bibinfo
  {author} {\bibfnamefont {B.~U.}\ \bibnamefont {Musch}},\ and\ \bibinfo
  {author} {\bibfnamefont {A.}~\bibnamefont {Sch\"afer}},\ }\href
  {https://doi.org/10.1103/PhysRevD.93.094515} {\bibfield  {journal} {\bibinfo
  {journal} {Phys. Rev. D}\ }\textbf {\bibinfo {volume} {93}},\ \bibinfo
  {pages} {094515} (\bibinfo {year} {2016})},\ \Eprint
  {https://arxiv.org/abs/1602.05525} {arXiv:1602.05525 [hep-lat]} \BibitemShut
  {NoStop}%
\bibitem [{\citenamefont {Huber}\ \emph
  {et~al.}(2008{\natexlab{b}})\citenamefont {Huber} \emph
  {et~al.}}]{JeffersonLab:2008jve}%
  \BibitemOpen
  \bibfield  {author} {\bibinfo {author} {\bibfnamefont {G.~M.}\ \bibnamefont
  {Huber}} \emph {et~al.} (\bibinfo {collaboration} {Jefferson Lab}),\ }\href
  {https://doi.org/10.1103/PhysRevC.78.045203} {\bibfield  {journal} {\bibinfo
  {journal} {Phys. Rev. C}\ }\textbf {\bibinfo {volume} {78}},\ \bibinfo
  {pages} {045203} (\bibinfo {year} {2008}{\natexlab{b}})},\ \Eprint
  {https://arxiv.org/abs/0809.3052} {arXiv:0809.3052 [nucl-ex]} \BibitemShut
  {NoStop}%
\bibitem [{\citenamefont {Cheng}(2019)}]{Cheng:2019ruz}%
  \BibitemOpen
  \bibfield  {author} {\bibinfo {author} {\bibfnamefont {S.}~\bibnamefont
  {Cheng}},\ }\href {https://doi.org/10.1103/PhysRevD.100.013007} {\bibfield
  {journal} {\bibinfo  {journal} {Phys. Rev. D}\ }\textbf {\bibinfo {volume}
  {100}},\ \bibinfo {pages} {013007} (\bibinfo {year} {2019})},\ \Eprint
  {https://arxiv.org/abs/1905.05059} {arXiv:1905.05059 [hep-ph]} \BibitemShut
  {NoStop}%
\bibitem [{\citenamefont {Chai}\ \emph {et~al.}(2023)\citenamefont {Chai},
  \citenamefont {Cheng},\ and\ \citenamefont {Hua}}]{Chai:2022srx}%
  \BibitemOpen
  \bibfield  {author} {\bibinfo {author} {\bibfnamefont {J.}~\bibnamefont
  {Chai}}, \bibinfo {author} {\bibfnamefont {S.}~\bibnamefont {Cheng}},\ and\
  \bibinfo {author} {\bibfnamefont {J.}~\bibnamefont {Hua}},\ }\href
  {https://doi.org/10.1140/epjc/s10052-023-11725-2} {\bibfield  {journal}
  {\bibinfo  {journal} {Eur. Phys. J. C}\ }\textbf {\bibinfo {volume} {83}},\
  \bibinfo {pages} {556} (\bibinfo {year} {2023})},\ \Eprint
  {https://arxiv.org/abs/2209.13312} {arXiv:2209.13312 [hep-ph]} \BibitemShut
  {NoStop}%
\bibitem [{\citenamefont {Chai}\ \emph {et~al.}()\citenamefont {Chai},
  \citenamefont {Cheng},\ and\ \citenamefont {Fang}}]{New_kaon_pQCD}%
  \BibitemOpen
  \bibfield  {author} {\bibinfo {author} {\bibfnamefont {J.}~\bibnamefont
  {Chai}}, \bibinfo {author} {\bibfnamefont {S.}~\bibnamefont {Cheng}},\ and\
  \bibinfo {author} {\bibfnamefont {Z.}~\bibnamefont {Fang}},\ }\href@noop {}
  {\bibinfo  {journal} {The transition and electromagnetic form factors of the
  pseudoscalar mesons, in preparation}\ }\BibitemShut {NoStop}%
\bibitem [{\citenamefont {Yao}\ \emph {et~al.}(2024)\citenamefont {Yao},
  \citenamefont {Binosi},\ and\ \citenamefont {Roberts}}]{Yao:2024drm}%
  \BibitemOpen
\bibfield  {journal} {  }\bibfield  {author} {\bibinfo {author} {\bibfnamefont
  {Z.-Q.}\ \bibnamefont {Yao}}, \bibinfo {author} {\bibfnamefont
  {D.}~\bibnamefont {Binosi}},\ and\ \bibinfo {author} {\bibfnamefont {C.~D.}\
  \bibnamefont {Roberts}},\ }\href
  {https://doi.org/10.1016/j.physletb.2024.138823} {\bibfield  {journal}
  {\bibinfo  {journal} {Phys. Lett. B}\ }\textbf {\bibinfo {volume} {855}},\
  \bibinfo {pages} {138823} (\bibinfo {year} {2024})},\ \Eprint
  {https://arxiv.org/abs/2405.04681} {arXiv:2405.04681 [hep-ph]} \BibitemShut
  {NoStop}%
\bibitem [{\citenamefont {Ydrefors}\ \emph {et~al.}(2021)\citenamefont
  {Ydrefors}, \citenamefont {de~Paula}, \citenamefont {Nogueira}, \citenamefont
  {Frederico},\ and\ \citenamefont {Salm\'e}}]{Ydrefors:2021dwa}%
  \BibitemOpen
  \bibfield  {author} {\bibinfo {author} {\bibfnamefont {E.}~\bibnamefont
  {Ydrefors}}, \bibinfo {author} {\bibfnamefont {W.}~\bibnamefont {de~Paula}},
  \bibinfo {author} {\bibfnamefont {J.~H.~A.}\ \bibnamefont {Nogueira}},
  \bibinfo {author} {\bibfnamefont {T.}~\bibnamefont {Frederico}},\ and\
  \bibinfo {author} {\bibfnamefont {G.}~\bibnamefont {Salm\'e}},\ }\href
  {https://doi.org/10.1016/j.physletb.2021.136494} {\bibfield  {journal}
  {\bibinfo  {journal} {Phys. Lett. B}\ }\textbf {\bibinfo {volume} {820}},\
  \bibinfo {pages} {136494} (\bibinfo {year} {2021})},\ \Eprint
  {https://arxiv.org/abs/2106.10018} {arXiv:2106.10018 [hep-ph]} \BibitemShut
  {NoStop}%
\bibitem [{\citenamefont {Jia}\ and\ \citenamefont
  {Clo\"et}(2024)}]{Jia:2024dfl}%
  \BibitemOpen
  \bibfield  {author} {\bibinfo {author} {\bibfnamefont {S.}~\bibnamefont
  {Jia}}\ and\ \bibinfo {author} {\bibfnamefont {I.}~\bibnamefont {Clo\"et}},\
  }\href@noop {} {\  (\bibinfo {year} {2024})},\ \Eprint
  {https://arxiv.org/abs/2402.00285} {arXiv:2402.00285 [hep-ph]} \BibitemShut
  {NoStop}%
\bibitem [{\citenamefont {Aoki}\ \emph {et~al.}(2022)\citenamefont {Aoki} \emph
  {et~al.}}]{Aoki:2021kgd}%
  \BibitemOpen
  \bibfield  {author} {\bibinfo {author} {\bibfnamefont {Y.}~\bibnamefont
  {Aoki}} \emph {et~al.} (\bibinfo {collaboration} {Flavour Lattice Averaging
  Group (FLAG)}),\ }\href {https://doi.org/10.1140/epjc/s10052-022-10536-1}
  {\bibfield  {journal} {\bibinfo  {journal} {Eur. Phys. J. C}\ }\textbf
  {\bibinfo {volume} {82}},\ \bibinfo {pages} {869} (\bibinfo {year} {2022})},\
  \Eprint {https://arxiv.org/abs/2111.09849} {arXiv:2111.09849 [hep-lat]}
  \BibitemShut {NoStop}%
\bibitem [{\citenamefont {Blum}\ \emph {et~al.}(2016)\citenamefont {Blum} \emph
  {et~al.}}]{RBC:2014ntl}%
  \BibitemOpen
  \bibfield  {author} {\bibinfo {author} {\bibfnamefont {T.}~\bibnamefont
  {Blum}} \emph {et~al.} (\bibinfo {collaboration} {RBC, UKQCD}),\ }\href
  {https://doi.org/10.1103/PhysRevD.93.074505} {\bibfield  {journal} {\bibinfo
  {journal} {Phys. Rev. D}\ }\textbf {\bibinfo {volume} {93}},\ \bibinfo
  {pages} {074505} (\bibinfo {year} {2016})},\ \Eprint
  {https://arxiv.org/abs/1411.7017} {arXiv:1411.7017 [hep-lat]} \BibitemShut
  {NoStop}%
\bibitem [{\citenamefont {Follana}\ \emph {et~al.}(2008)\citenamefont
  {Follana}, \citenamefont {Davies}, \citenamefont {Lepage},\ and\
  \citenamefont {Shigemitsu}}]{Follana:2007uv}%
  \BibitemOpen
  \bibfield  {author} {\bibinfo {author} {\bibfnamefont {E.}~\bibnamefont
  {Follana}}, \bibinfo {author} {\bibfnamefont {C.~T.~H.}\ \bibnamefont
  {Davies}}, \bibinfo {author} {\bibfnamefont {G.~P.}\ \bibnamefont {Lepage}},\
  and\ \bibinfo {author} {\bibfnamefont {J.}~\bibnamefont {Shigemitsu}}
  (\bibinfo {collaboration} {HPQCD, UKQCD}),\ }\href
  {https://doi.org/10.1103/PhysRevLett.100.062002} {\bibfield  {journal}
  {\bibinfo  {journal} {Phys. Rev. Lett.}\ }\textbf {\bibinfo {volume} {100}},\
  \bibinfo {pages} {062002} (\bibinfo {year} {2008})},\ \Eprint
  {https://arxiv.org/abs/0706.1726} {arXiv:0706.1726 [hep-lat]} \BibitemShut
  {NoStop}%
\bibitem [{\citenamefont {Bazavov}\ \emph {et~al.}(2010)\citenamefont {Bazavov}
  \emph {et~al.}}]{MILC:2010hzw}%
  \BibitemOpen
  \bibfield  {author} {\bibinfo {author} {\bibfnamefont {A.}~\bibnamefont
  {Bazavov}} \emph {et~al.} (\bibinfo {collaboration} {MILC}),\ }\href
  {https://doi.org/10.22323/1.105.0074} {\bibfield  {journal} {\bibinfo
  {journal} {PoS}\ }\textbf {\bibinfo {volume} {LATTICE2010}},\ \bibinfo
  {pages} {074} (\bibinfo {year} {2010})},\ \Eprint
  {https://arxiv.org/abs/1012.0868} {arXiv:1012.0868 [hep-lat]} \BibitemShut
  {NoStop}%
\bibitem [{\citenamefont {Efremov}\ and\ \citenamefont
  {Radyushkin}(1980)}]{Efremov:1979qk}%
  \BibitemOpen
  \bibfield  {author} {\bibinfo {author} {\bibfnamefont {A.~V.}\ \bibnamefont
  {Efremov}}\ and\ \bibinfo {author} {\bibfnamefont {A.~V.}\ \bibnamefont
  {Radyushkin}},\ }\href {https://doi.org/10.1016/0370-2693(80)90869-2}
  {\bibfield  {journal} {\bibinfo  {journal} {Phys. Lett. B}\ }\textbf
  {\bibinfo {volume} {94}},\ \bibinfo {pages} {245} (\bibinfo {year}
  {1980})}\BibitemShut {NoStop}%
\bibitem [{\citenamefont {Field}\ \emph {et~al.}(1981)\citenamefont {Field},
  \citenamefont {Gupta}, \citenamefont {Otto},\ and\ \citenamefont
  {Chang}}]{Field:1981wx}%
  \BibitemOpen
  \bibfield  {author} {\bibinfo {author} {\bibfnamefont {R.~D.}\ \bibnamefont
  {Field}}, \bibinfo {author} {\bibfnamefont {R.}~\bibnamefont {Gupta}},
  \bibinfo {author} {\bibfnamefont {S.}~\bibnamefont {Otto}},\ and\ \bibinfo
  {author} {\bibfnamefont {L.}~\bibnamefont {Chang}},\ }\href
  {https://doi.org/10.1016/0550-3213(81)90022-5} {\bibfield  {journal}
  {\bibinfo  {journal} {Nucl. Phys. B}\ }\textbf {\bibinfo {volume} {186}},\
  \bibinfo {pages} {429} (\bibinfo {year} {1981})}\BibitemShut {NoStop}%
\bibitem [{\citenamefont {Dittes}\ and\ \citenamefont
  {Radyushkin}(1981)}]{Dittes:1981aw}%
  \BibitemOpen
  \bibfield  {author} {\bibinfo {author} {\bibfnamefont {F.~M.}\ \bibnamefont
  {Dittes}}\ and\ \bibinfo {author} {\bibfnamefont {A.~V.}\ \bibnamefont
  {Radyushkin}},\ }\href@noop {} {\bibfield  {journal} {\bibinfo  {journal}
  {Sov. J. Nucl. Phys.}\ }\textbf {\bibinfo {volume} {34}},\ \bibinfo {pages}
  {293} (\bibinfo {year} {1981})}\BibitemShut {NoStop}%
\bibitem [{\citenamefont {Sarmadi}(1984)}]{Sarmadi:1982yg}%
  \BibitemOpen
  \bibfield  {author} {\bibinfo {author} {\bibfnamefont {M.~H.}\ \bibnamefont
  {Sarmadi}},\ }\href {https://doi.org/10.1016/0370-2693(84)91504-1} {\bibfield
   {journal} {\bibinfo  {journal} {Phys. Lett. B}\ }\textbf {\bibinfo {volume}
  {143}},\ \bibinfo {pages} {471} (\bibinfo {year} {1984})}\BibitemShut
  {NoStop}%
\bibitem [{\citenamefont {Braaten}\ and\ \citenamefont
  {Tse}(1987)}]{Braaten:1987yy}%
  \BibitemOpen
  \bibfield  {author} {\bibinfo {author} {\bibfnamefont {E.}~\bibnamefont
  {Braaten}}\ and\ \bibinfo {author} {\bibfnamefont {S.-M.}\ \bibnamefont
  {Tse}},\ }\href {https://doi.org/10.1103/PhysRevD.35.2255} {\bibfield
  {journal} {\bibinfo  {journal} {Phys. Rev. D}\ }\textbf {\bibinfo {volume}
  {35}},\ \bibinfo {pages} {2255} (\bibinfo {year} {1987})}\BibitemShut
  {NoStop}%
\bibitem [{\citenamefont {Nandi}\ and\ \citenamefont
  {Li}(2007)}]{Nandi:2007qx}%
  \BibitemOpen
  \bibfield  {author} {\bibinfo {author} {\bibfnamefont {S.}~\bibnamefont
  {Nandi}}\ and\ \bibinfo {author} {\bibfnamefont {H.-n.}\ \bibnamefont {Li}},\
  }\href {https://doi.org/10.1103/PhysRevD.76.034008} {\bibfield  {journal}
  {\bibinfo  {journal} {Phys. Rev. D}\ }\textbf {\bibinfo {volume} {76}},\
  \bibinfo {pages} {034008} (\bibinfo {year} {2007})},\ \Eprint
  {https://arxiv.org/abs/0704.3790} {arXiv:0704.3790 [hep-ph]} \BibitemShut
  {NoStop}%
\bibitem [{\citenamefont {Li}\ \emph {et~al.}(2011)\citenamefont {Li},
  \citenamefont {Shen}, \citenamefont {Wang},\ and\ \citenamefont
  {Zou}}]{Li:2010nn}%
  \BibitemOpen
  \bibfield  {author} {\bibinfo {author} {\bibfnamefont {H.-n.}\ \bibnamefont
  {Li}}, \bibinfo {author} {\bibfnamefont {Y.-L.}\ \bibnamefont {Shen}},
  \bibinfo {author} {\bibfnamefont {Y.-M.}\ \bibnamefont {Wang}},\ and\
  \bibinfo {author} {\bibfnamefont {H.}~\bibnamefont {Zou}},\ }\href
  {https://doi.org/10.1103/PhysRevD.83.054029} {\bibfield  {journal} {\bibinfo
  {journal} {Phys. Rev. D}\ }\textbf {\bibinfo {volume} {83}},\ \bibinfo
  {pages} {054029} (\bibinfo {year} {2011})},\ \Eprint
  {https://arxiv.org/abs/1012.4098} {arXiv:1012.4098 [hep-ph]} \BibitemShut
  {NoStop}%
\bibitem [{\citenamefont {Raha}\ and\ \citenamefont
  {Aste}(2009)}]{Raha:2008ve}%
  \BibitemOpen
  \bibfield  {author} {\bibinfo {author} {\bibfnamefont {U.}~\bibnamefont
  {Raha}}\ and\ \bibinfo {author} {\bibfnamefont {A.}~\bibnamefont {Aste}},\
  }\href {https://doi.org/10.1103/PhysRevD.79.034015} {\bibfield  {journal}
  {\bibinfo  {journal} {Phys. Rev. D}\ }\textbf {\bibinfo {volume} {79}},\
  \bibinfo {pages} {034015} (\bibinfo {year} {2009})},\ \Eprint
  {https://arxiv.org/abs/0809.1359} {arXiv:0809.1359 [hep-ph]} \BibitemShut
  {NoStop}%
\bibitem [{\citenamefont {Geshkenbein}\ and\ \citenamefont
  {Terentev}(1982)}]{Geshkenbein:1982zs}%
  \BibitemOpen
  \bibfield  {author} {\bibinfo {author} {\bibfnamefont {B.~V.}\ \bibnamefont
  {Geshkenbein}}\ and\ \bibinfo {author} {\bibfnamefont {M.~V.}\ \bibnamefont
  {Terentev}},\ }\href {https://doi.org/10.1016/0370-2693(82)90555-X}
  {\bibfield  {journal} {\bibinfo  {journal} {Phys. Lett. B}\ }\textbf
  {\bibinfo {volume} {117}},\ \bibinfo {pages} {243} (\bibinfo {year}
  {1982})}\BibitemShut {NoStop}%
\bibitem [{\citenamefont {Cao}\ \emph {et~al.}(1999)\citenamefont {Cao},
  \citenamefont {Dai},\ and\ \citenamefont {Huang}}]{Cao:1997st}%
  \BibitemOpen
  \bibfield  {author} {\bibinfo {author} {\bibfnamefont {F.-g.}\ \bibnamefont
  {Cao}}, \bibinfo {author} {\bibfnamefont {Y.-b.}\ \bibnamefont {Dai}},\ and\
  \bibinfo {author} {\bibfnamefont {C.-s.}\ \bibnamefont {Huang}},\ }\href
  {https://doi.org/10.1007/s100520050650} {\bibfield  {journal} {\bibinfo
  {journal} {Eur. Phys. J. C}\ }\textbf {\bibinfo {volume} {11}},\ \bibinfo
  {pages} {501} (\bibinfo {year} {1999})},\ \Eprint
  {https://arxiv.org/abs/hep-ph/9711203} {arXiv:hep-ph/9711203} \BibitemShut
  {NoStop}%
\bibitem [{\citenamefont {Braun}\ \emph {et~al.}(2000)\citenamefont {Braun},
  \citenamefont {Khodjamirian},\ and\ \citenamefont {Maul}}]{Braun:1999uj}%
  \BibitemOpen
  \bibfield  {author} {\bibinfo {author} {\bibfnamefont {V.~M.}\ \bibnamefont
  {Braun}}, \bibinfo {author} {\bibfnamefont {A.}~\bibnamefont
  {Khodjamirian}},\ and\ \bibinfo {author} {\bibfnamefont {M.}~\bibnamefont
  {Maul}},\ }\href {https://doi.org/10.1103/PhysRevD.61.073004} {\bibfield
  {journal} {\bibinfo  {journal} {Phys. Rev. D}\ }\textbf {\bibinfo {volume}
  {61}},\ \bibinfo {pages} {073004} (\bibinfo {year} {2000})},\ \Eprint
  {https://arxiv.org/abs/hep-ph/9907495} {arXiv:hep-ph/9907495} \BibitemShut
  {NoStop}%
\bibitem [{\citenamefont {Bijnens}\ and\ \citenamefont
  {Khodjamirian}(2002)}]{Bijnens:2002mg}%
  \BibitemOpen
  \bibfield  {author} {\bibinfo {author} {\bibfnamefont {J.}~\bibnamefont
  {Bijnens}}\ and\ \bibinfo {author} {\bibfnamefont {A.}~\bibnamefont
  {Khodjamirian}},\ }\href {https://doi.org/10.1140/epjc/s2002-01042-1}
  {\bibfield  {journal} {\bibinfo  {journal} {Eur. Phys. J. C}\ }\textbf
  {\bibinfo {volume} {26}},\ \bibinfo {pages} {67} (\bibinfo {year} {2002})},\
  \Eprint {https://arxiv.org/abs/hep-ph/0206252} {arXiv:hep-ph/0206252}
  \BibitemShut {NoStop}%
\bibitem [{\citenamefont {Wei}\ and\ \citenamefont {Yang}(2003)}]{Wei:2003fm}%
  \BibitemOpen
  \bibfield  {author} {\bibinfo {author} {\bibfnamefont {Z.-T.}\ \bibnamefont
  {Wei}}\ and\ \bibinfo {author} {\bibfnamefont {M.-Z.}\ \bibnamefont {Yang}},\
  }\href {https://doi.org/10.1103/PhysRevD.67.094013} {\bibfield  {journal}
  {\bibinfo  {journal} {Phys. Rev. D}\ }\textbf {\bibinfo {volume} {67}},\
  \bibinfo {pages} {094013} (\bibinfo {year} {2003})}\BibitemShut {NoStop}%
\bibitem [{\citenamefont {Huang}\ and\ \citenamefont
  {Wu}(2004)}]{Huang:2004su}%
  \BibitemOpen
  \bibfield  {author} {\bibinfo {author} {\bibfnamefont {T.}~\bibnamefont
  {Huang}}\ and\ \bibinfo {author} {\bibfnamefont {X.-G.}\ \bibnamefont {Wu}},\
  }\href {https://doi.org/10.1103/PhysRevD.70.093013} {\bibfield  {journal}
  {\bibinfo  {journal} {Phys. Rev. D}\ }\textbf {\bibinfo {volume} {70}},\
  \bibinfo {pages} {093013} (\bibinfo {year} {2004})},\ \Eprint
  {https://arxiv.org/abs/hep-ph/0408252} {arXiv:hep-ph/0408252} \BibitemShut
  {NoStop}%
\bibitem [{\citenamefont {Braun}\ \emph {et~al.}(2017)\citenamefont {Braun},
  \citenamefont {Manashov}, \citenamefont {Moch},\ and\ \citenamefont
  {Strohmaier}}]{Braun:2017cih}%
  \BibitemOpen
  \bibfield  {author} {\bibinfo {author} {\bibfnamefont {V.~M.}\ \bibnamefont
  {Braun}}, \bibinfo {author} {\bibfnamefont {A.~N.}\ \bibnamefont {Manashov}},
  \bibinfo {author} {\bibfnamefont {S.}~\bibnamefont {Moch}},\ and\ \bibinfo
  {author} {\bibfnamefont {M.}~\bibnamefont {Strohmaier}},\ }\href
  {https://doi.org/10.1007/JHEP06(2017)037} {\bibfield  {journal} {\bibinfo
  {journal} {JHEP}\ }\textbf {\bibinfo {volume} {06}},\ \bibinfo {pages}
  {037}},\ \Eprint {https://arxiv.org/abs/1703.09532} {arXiv:1703.09532
  [hep-ph]} \BibitemShut {NoStop}%
\bibitem [{\citenamefont {Gao}\ \emph {et~al.}(2017)\citenamefont {Gao},
  \citenamefont {Chang}, \citenamefont {Liu}, \citenamefont {Roberts},\ and\
  \citenamefont {Tandy}}]{Gao:2017mmp}%
  \BibitemOpen
  \bibfield  {author} {\bibinfo {author} {\bibfnamefont {F.}~\bibnamefont
  {Gao}}, \bibinfo {author} {\bibfnamefont {L.}~\bibnamefont {Chang}}, \bibinfo
  {author} {\bibfnamefont {Y.-X.}\ \bibnamefont {Liu}}, \bibinfo {author}
  {\bibfnamefont {C.~D.}\ \bibnamefont {Roberts}},\ and\ \bibinfo {author}
  {\bibfnamefont {P.~C.}\ \bibnamefont {Tandy}},\ }\href
  {https://doi.org/10.1103/PhysRevD.96.034024} {\bibfield  {journal} {\bibinfo
  {journal} {Phys. Rev. D}\ }\textbf {\bibinfo {volume} {96}},\ \bibinfo
  {pages} {034024} (\bibinfo {year} {2017})},\ \Eprint
  {https://arxiv.org/abs/1703.04875} {arXiv:1703.04875 [nucl-th]} \BibitemShut
  {NoStop}%
\bibitem [{\citenamefont {O'Connell}\ \emph {et~al.}(1995)\citenamefont
  {O'Connell}, \citenamefont {Pearce}, \citenamefont {Thomas},\ and\
  \citenamefont {Williams}}]{OConnell:1995fwv}%
  \BibitemOpen
  \bibfield  {author} {\bibinfo {author} {\bibfnamefont {H.~B.}\ \bibnamefont
  {O'Connell}}, \bibinfo {author} {\bibfnamefont {B.~C.}\ \bibnamefont
  {Pearce}}, \bibinfo {author} {\bibfnamefont {A.~W.}\ \bibnamefont {Thomas}},\
  and\ \bibinfo {author} {\bibfnamefont {A.~G.}\ \bibnamefont {Williams}},\
  }\href {https://doi.org/10.1016/0370-2693(95)00642-X} {\bibfield  {journal}
  {\bibinfo  {journal} {Phys. Lett. B}\ }\textbf {\bibinfo {volume} {354}},\
  \bibinfo {pages} {14} (\bibinfo {year} {1995})},\ \Eprint
  {https://arxiv.org/abs/hep-ph/9503332} {arXiv:hep-ph/9503332} \BibitemShut
  {NoStop}%
\bibitem [{\citenamefont {Klingl}\ \emph {et~al.}(1996)\citenamefont {Klingl},
  \citenamefont {Kaiser},\ and\ \citenamefont {Weise}}]{Klingl:1996by}%
  \BibitemOpen
  \bibfield  {author} {\bibinfo {author} {\bibfnamefont {F.}~\bibnamefont
  {Klingl}}, \bibinfo {author} {\bibfnamefont {N.}~\bibnamefont {Kaiser}},\
  and\ \bibinfo {author} {\bibfnamefont {W.}~\bibnamefont {Weise}},\ }\href
  {https://doi.org/10.1007/s002180050167} {\bibfield  {journal} {\bibinfo
  {journal} {Z. Phys. A}\ }\textbf {\bibinfo {volume} {356}},\ \bibinfo {pages}
  {193} (\bibinfo {year} {1996})},\ \Eprint
  {https://arxiv.org/abs/hep-ph/9607431} {arXiv:hep-ph/9607431} \BibitemShut
  {NoStop}%
\bibitem [{\citenamefont {Ivashyn}\ and\ \citenamefont
  {Korchin}(2007)}]{Ivashyn:2006gf}%
  \BibitemOpen
  \bibfield  {author} {\bibinfo {author} {\bibfnamefont {S.~A.}\ \bibnamefont
  {Ivashyn}}\ and\ \bibinfo {author} {\bibfnamefont {A.~Y.}\ \bibnamefont
  {Korchin}},\ }\href {https://doi.org/10.1140/epjc/s10052-006-0167-5}
  {\bibfield  {journal} {\bibinfo  {journal} {Eur. Phys. J. C}\ }\textbf
  {\bibinfo {volume} {49}},\ \bibinfo {pages} {697} (\bibinfo {year} {2007})},\
  \Eprint {https://arxiv.org/abs/hep-ph/0611039} {arXiv:hep-ph/0611039}
  \BibitemShut {NoStop}%
\bibitem [{\citenamefont {Stamen}\ \emph {et~al.}(2022)\citenamefont {Stamen},
  \citenamefont {Hariharan}, \citenamefont {Hoferichter}, \citenamefont
  {Kubis},\ and\ \citenamefont {Stoffer}}]{Stamen:2022uqh}%
  \BibitemOpen
  \bibfield  {author} {\bibinfo {author} {\bibfnamefont {D.}~\bibnamefont
  {Stamen}}, \bibinfo {author} {\bibfnamefont {D.}~\bibnamefont {Hariharan}},
  \bibinfo {author} {\bibfnamefont {M.}~\bibnamefont {Hoferichter}}, \bibinfo
  {author} {\bibfnamefont {B.}~\bibnamefont {Kubis}},\ and\ \bibinfo {author}
  {\bibfnamefont {P.}~\bibnamefont {Stoffer}},\ }\href
  {https://doi.org/10.1140/epjc/s10052-022-10348-3} {\bibfield  {journal}
  {\bibinfo  {journal} {Eur. Phys. J. C}\ }\textbf {\bibinfo {volume} {82}},\
  \bibinfo {pages} {432} (\bibinfo {year} {2022})},\ \Eprint
  {https://arxiv.org/abs/2202.11106} {arXiv:2202.11106 [hep-ph]} \BibitemShut
  {NoStop}%
\bibitem [{\citenamefont {Workman}\ \emph {et~al.}(2022)\citenamefont {Workman}
  \emph {et~al.}}]{Workman:2022ynf}%
  \BibitemOpen
  \bibfield  {author} {\bibinfo {author} {\bibfnamefont {R.~L.}\ \bibnamefont
  {Workman}} \emph {et~al.} (\bibinfo {collaboration} {Particle Data Group}),\
  }\href {https://doi.org/10.1093/ptep/ptac097} {\bibfield  {journal} {\bibinfo
   {journal} {PTEP}\ }\textbf {\bibinfo {volume} {2022}},\ \bibinfo {pages}
  {083C01} (\bibinfo {year} {2022})}\BibitemShut {NoStop}%
\bibitem [{\citenamefont {Bhattacharya}\ \emph {et~al.}(2023)\citenamefont
  {Bhattacharya}, \citenamefont {Cichy}, \citenamefont {Constantinou},
  \citenamefont {Gao}, \citenamefont {Metz}, \citenamefont {Miller},
  \citenamefont {Mukherjee}, \citenamefont {Petreczky}, \citenamefont
  {Steffens},\ and\ \citenamefont {Zhao}}]{Bhattacharya:2023ays}%
  \BibitemOpen
  \bibfield  {author} {\bibinfo {author} {\bibfnamefont {S.}~\bibnamefont
  {Bhattacharya}}, \bibinfo {author} {\bibfnamefont {K.}~\bibnamefont {Cichy}},
  \bibinfo {author} {\bibfnamefont {M.}~\bibnamefont {Constantinou}}, \bibinfo
  {author} {\bibfnamefont {X.}~\bibnamefont {Gao}}, \bibinfo {author}
  {\bibfnamefont {A.}~\bibnamefont {Metz}}, \bibinfo {author} {\bibfnamefont
  {J.}~\bibnamefont {Miller}}, \bibinfo {author} {\bibfnamefont
  {S.}~\bibnamefont {Mukherjee}}, \bibinfo {author} {\bibfnamefont
  {P.}~\bibnamefont {Petreczky}}, \bibinfo {author} {\bibfnamefont
  {F.}~\bibnamefont {Steffens}},\ and\ \bibinfo {author} {\bibfnamefont
  {Y.}~\bibnamefont {Zhao}},\ }\href
  {https://doi.org/10.1103/PhysRevD.108.014507} {\bibfield  {journal} {\bibinfo
   {journal} {Phys. Rev. D}\ }\textbf {\bibinfo {volume} {108}},\ \bibinfo
  {pages} {014507} (\bibinfo {year} {2023})},\ \Eprint
  {https://arxiv.org/abs/2305.11117} {arXiv:2305.11117 [hep-lat]} \BibitemShut
  {NoStop}%
\bibitem [{\citenamefont {Hutauruk}\ \emph {et~al.}(2016)\citenamefont
  {Hutauruk}, \citenamefont {Cloet},\ and\ \citenamefont
  {Thomas}}]{Hutauruk:2016sug}%
  \BibitemOpen
  \bibfield  {author} {\bibinfo {author} {\bibfnamefont {P.~T.~P.}\
  \bibnamefont {Hutauruk}}, \bibinfo {author} {\bibfnamefont {I.~C.}\
  \bibnamefont {Cloet}},\ and\ \bibinfo {author} {\bibfnamefont {A.~W.}\
  \bibnamefont {Thomas}},\ }\href {https://doi.org/10.1103/PhysRevC.94.035201}
  {\bibfield  {journal} {\bibinfo  {journal} {Phys. Rev. C}\ }\textbf {\bibinfo
  {volume} {94}},\ \bibinfo {pages} {035201} (\bibinfo {year} {2016})},\
  \Eprint {https://arxiv.org/abs/1604.02853} {arXiv:1604.02853 [nucl-th]}
  \BibitemShut {NoStop}%
\end{thebibliography}%

\parskip=6pt
\newpage
\begin{widetext}

% \appendix
\section*{Supplemental Materials}

\section{The electromagnetic form factors of pion and kaon}\label{sec:FFvalue}

The values of the momenta of the initial and final meson states
used in this study are summarized in \tb{setup}. On a finite
lattice, the initial and final momenta are given 
as $\mathbf{P}^{i,f}=2 \pi/(L_s a)\mathbf{n}^{i,f}_{\mathbf{P}}$,
with $a$ being the lattice spacing and $L_s$ being the spatial grid size. The different choices of $\mathbf{n}^{i,f}$
are listed in \tb{setup}. For each choice of $\mathbf{n}^{i,f}$,
we list the corresponding quark boost momentum along the $z$
direction $k_3^{i,f}=2 \pi/(L_s a) n_{k_3}^{i,f}$ for the
initial and final states, as well as the momentum
transfer $Q^2$. Additionally, we give the number of the exact and sloppy AMA samples.

The lattice results of the pion and kaon EMFF derived in this work are summarized in \tb{EMFFpion} and \tb{EMFFkaon}, respectively, which are also shown in \fig{FF_K_pi} as a function of $Q^2$. For the case of pion, we also include the results from BNL21~\cite{Gao:2021xsm} using the same lattice setup of the $a$ = 0.076 fm ensemble. For the case of kaon, we have two different lattice ensembles. Moreover, we also calculated
the form factors of the valence quark. The results are shown in 
\fig{FF_K_quark} along with the corresponding first-order spline interpolations.
Using the spline interpolations, we can perform the numerical Fourier transformation of the valence quark
form factors and obtain the distribution of $u$ quark inside the pion and kaon, as well
as the distribution of $s$ anti-quark inside kaon in the transverse plane, i.e., distribution
of valence quark in $b_{\perp}$. Here $b_{\perp}$ is the conjugate variable to $Q$~\cite{Bhattacharya:2023ays}, and we used $Q^2$ up to 10 $\rm GeV^2$ for both pion and kaon for consistency.
These distributions in the transverse plane are shown in the right panel of \fig{FF_K_quark}.
We see that the valence $u$ quark $b_{\perp}$ distributions in pion and kaon are about the same,
while the distribution of $s$ anti-quark is narrower.

% \squeezetable
\begin{table}[t!]
\centering
\rowheight{1.4} % Adjust the row height multiplier
\caption{\label{tb:setup} The values of the meson initial and final momenta used in this work and the numbers of the samples used in the AMA method are shown. The initial and final momenta and their corresponding boost parameters in the $z$-direction in units of $2\pi/(L_s a)$ are denoted by $\mathbf{n}^{i,f}_{\mathbf{P}}$ and $n^{i,f}_{k_3}$, respectively. The values of the momentum transfer $Q^2$ are given in physical units. Additionally, we provide the exact counts for both exact (\#ex) and sloppy (\#sl) inversion samples per configuration.} 
\begin{tabular}{C{1.5cm}|C{1.5cm}|C{3.6cm}C{1cm}C{3.6cm}C{1cm}C{2cm}C{2cm}}
\hline
\hline
Meson & $a$[fm] & $\mathbf{n}^f_{\mathbf{P}}=(n^f_{P_1}, n^f_{P_2}, n^f_{P_3})$ & $n^f_{k_3}$ & $\mathbf{n}^i_{\mathbf{P}}=(n^i_{P_1}, n^i_{P_2}, n^i_{P_3})$ & $n^i_{k_3}$ & $Q^2$[GeV$^2$] & (\#ex, \#sl) \\
% \colrule
\hline
\multirow{13}{*}{Pion} & \multirow{13}{*}{$0.076$} & \multirow{4}{*}{(0, 0, -3)} & \multirow{4}{*}{-2} & (0,0,2) & \multirow{4}{*}{2} & 1.56 & \multirow{4}{*}{(3, 96)} \\
&&&& (0,0,3) && 2.34 & \\
&&&& (0,0,4) && 3.12 & \\
&&&& (2,0,3) && 2.58 & \\
\cline{3-8}
&& \multirow{5}{*}{(0, 0, -5)} & \multirow{5}{*}{-4} & (0,0,3) & \multirow{5}{*}{4} & 3.90 & \multirow{5}{*}{(7, 224)} \\
&&&& (0,0,4) && 5.20 & \\
&&&& (0,0,5) && 6.50 & \\
&&&& (2,0,4) && 5.50 & \\
&&&& (2,0,5) && 6.75 & \\
\cline{3-8}
&& \multirow{4}{*}{(0, 0, -6)} & \multirow{4}{*}{-5} & (0,0,5) & \multirow{4}{*}{5} & 7.80 & \multirow{4}{*}{(18, 576)} \\
&&&& (0,0,6) && 9.35 & \\
&&&& (2,0,5) && 8.10 & \\
&&&& (2,0,6) && 9.61 & \\
\hline
\multirow{23}{*}{Kaon} & \multirow{21}{*}{0.076} & \multirow{6}{*}{(0, 0, -1)} & \multirow{6}{*}{0} & (0,0,1) & \multirow{6}{*}{0} & {0.26} & \multirow{6}{*}{(1, 32)} \\
&&&& (1,0,-1) && 0.062 & \\
&&&& (1,0,0) && 0.13 & \\
&&&& (1,0,1) && 0.32 & \\
&&&& (1,1,0) && 0.19 & \\
&&&& (1,1,1) && 0.38 & \\
\cline{3-8}
&& \multirow{6}{*}{(0, 0, -3)} & \multirow{6}{*}{-2} & (0,0,-2) & \multirow{6}{*}{2} & 0.025 & \multirow{6}{*}{(5, 160)} \\
&&&& (0,0,1) && 0.91 & \\
&&&& (0,0,2) && 1.58 & \\
&&&& (0,0,3) && 2.34 & \\
&&&& (1,1,3) && 2.46 & \\
&&&& (2,2,3) && 2.80 & \\
\cline{3-8}
&& \multirow{5}{*}{(0, 0, -5)} & \multirow{5}{*}{-4} & (0,0,3) & \multirow{5}{*}{4} & 3.95 & \multirow{5}{*}{(8, 256)} \\
&&&& (0,0,4) && 5.21 & \\
&&&& (0,0,5) && 6.50 & \\
&&&& (1,1,5) && 6.62 & \\
&&&& (2,2,5) && 6.98 & \\
\cline{3-8}
&& \multirow{4}{*}{(0, 0, -6)} & \multirow{4}{*}{-6} & (0,0,5) & \multirow{4}{*}{6} & 7.80 & \multirow{4}{*}{(8, 256)} \\
&&&& (0,0,6) && 9.35 & \\
&&&& (1,1,6) && 9.48 & \\
&&&& (2,2,6) && 9.84 & \\
\cline{2-8}
& \multirow{7}{*}{0.04} & \multirow{4}{*}{(0, 0, -3)} & \multirow{4}{*}{-2} & (0,0,2) & \multirow{4}{*}{2} & 5.66 & \multirow{4}{*}{(4, 128)} \\
&&&& (0,0,3) && 8.44 & \\
&&&& (0,0,4) && 11.27 & \\
&&&& (0,0,5) && 14.13 & \\
\cline{3-8}
&& \multirow{3}{*}{(0, 0, -5)} & \multirow{3}{*}{-4} & (0,0,4) & \multirow{3}{*}{4} & 18.77 & \multirow{3}{*}{(8, 256)} \\
&&&& (0,0,5) && 23.45 & \\
&&&& (0,0,6) && 28.15 & \\
\hline
\hline
\end{tabular}
\end{table}

\begin{table}
\rowheight{1.4} % Adjust the row height multiplier
\centering
\begin{tabular}{c c c c c c c c c c c c}
\hline
\hline
$Q^2$ [$\rm GeV^2$]& 1.56 & 2.34 & 2.58 & 3.12 & 3.90 & 5.20\cr
\hline
$F_{\pi^+}$& 0.252(5) & 0.176(8) & 0.169(6) &0.142(22) &0.120(9) &0.096(11)\cr
\hline
\hline
5.50 & 6.50 & 6.75 & 7.80 & 8.10 & 9.35 & 9.61\cr
\hline
0.098(9) & 0.073(11) &0.068(11) &0.068(11) & 0.061(8) &0.056(10) &0.053(9)\cr
\hline
\hline
\end{tabular}
\caption{Pion electromagnetic form factors in this work from the $a$ = 0.076 fm lattice are shown.
}
\label{tb:EMFFpion}
\end{table}

\begin{table}
\rowheight{1.4} % Adjust the row height multiplier
\centering
\begin{tabular}{c c c c c c c c c c c c}
\hline
\hline
$Q^2$ [$\rm GeV^2$]& 0.025 & 0.062 & 0.13 & 0.19 & 0.26 & 0.32 & 0.38 & 0.91 & 1.58\cr
\hline
$F_{K^+}$& 0.967(14) & 0.906(2) & 0.828(1) &0.767(1) &0.710(1) &0.666(1) & 0.628(1) & 0.414(1) &0.283(1)\cr
\hline
\hline
2.34 & 2.46 & 2.80 & 3.95 & 5.21 & 5.60 & 6.50 & 6.62 & 6.98 & 7.80\cr
\hline
0.207(3) &0.200(1) &0.184(2) &0.135(2) &0.105(1) & 0.103(4) &0.088(1) &0.086(1) &0.082(2) &0.072(4)\cr
\hline
\hline
8.44 & 9.35 & 9.48 & 9.85 & 11.26 & 14.13 & 18.63 & 23.45 & 28.12\cr
\hline
0.075(4) & 0.063(3) &0.062(3) &0.056(4) &0.053(4) & 0.038(13) &0.033(13) &0.025(4) &0.019(3)\cr
\hline
\hline
\end{tabular}
\caption{Kaon electromagnetic form factors from the $a$ = 0.076 fm and $a$ = 0.04 fm lattices are shown.
}
\label{tb:EMFFkaon}
\end{table}

\begin{figure}
	\includegraphics[width=0.45\textwidth]{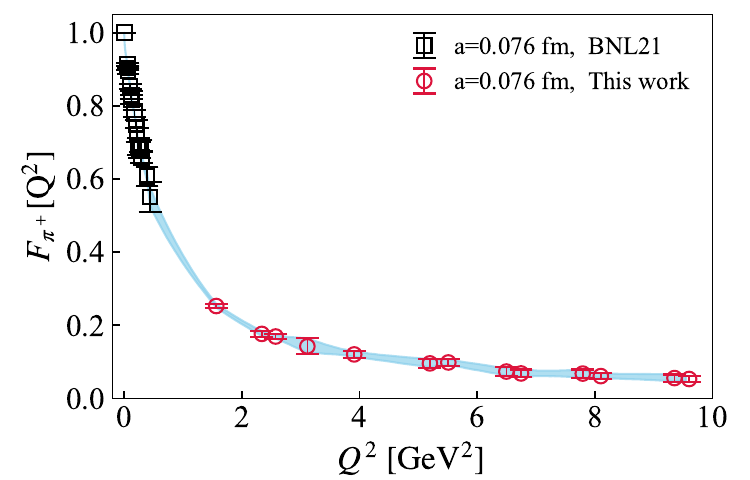}
	\includegraphics[width=0.45\textwidth]{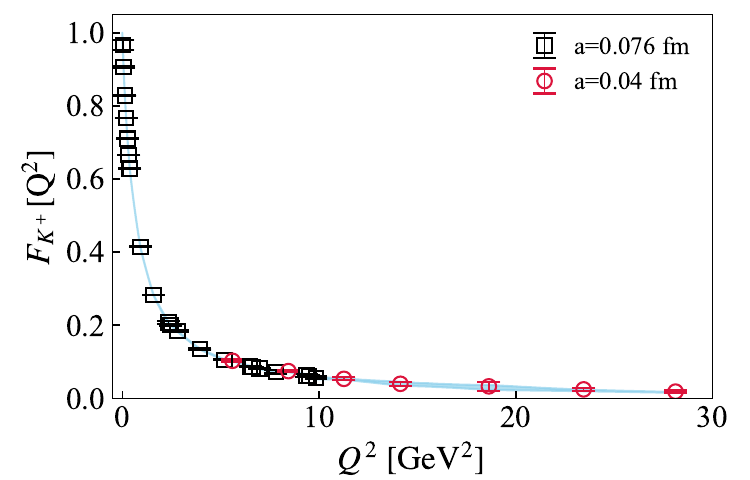}
	\caption{The EMFF of pion (left panel) and kaon (right panel) are shown as a function of $Q^2$. For the case of pion, we also include the results from BNL21~\cite{Gao:2021xsm} using the same lattice setup of the $a$ = 0.076 fm ensemble. For the case of kaon, we have two different ensembles. The bands are from spline interpolation of the first order.\label{fig:FF_K_pi}}
\end{figure}

\begin{figure}
    \includegraphics[width=0.45\textwidth]{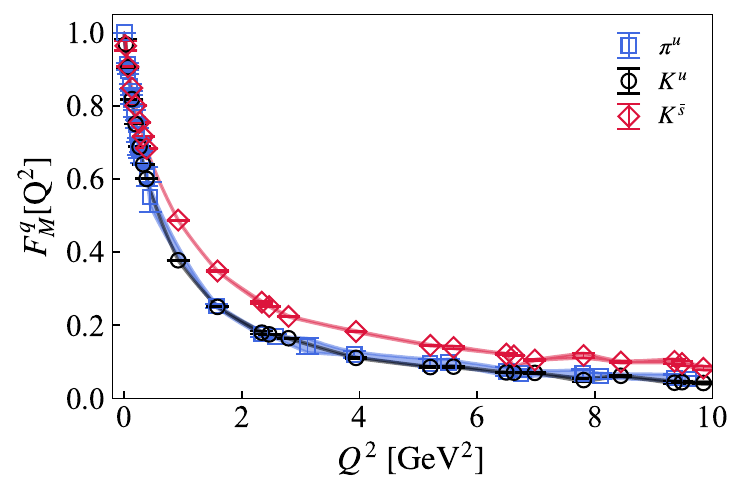}
    \includegraphics[width=0.45\textwidth]{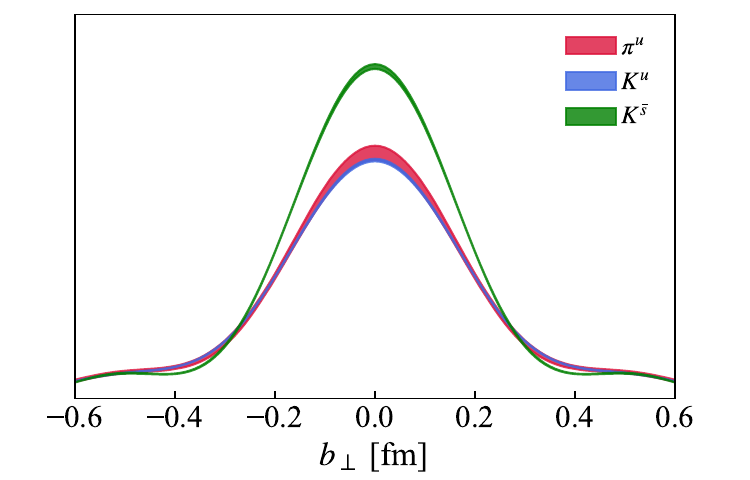}
	\caption{Left: The form factors of valence quarks in pion and kaon are shown as a function of $Q^2$. The bands are from spline interpolation of the first order. Right: The flavor distributions of the valence quarks in pion and kaon in the impact-parameter plane are shown as a function of transverse distance $b_\perp$.\label{fig:FF_K_quark}}
\end{figure}

\section{The bare matrix elements of kaon form factors}\label{sec:bm}

The analysis of the two-point and three-point function of the pion and kaon,
and the extraction of the bare matrix elements follows
the strategy laid out in \refcite{Gao:2021xsm, Gao:2022iex}. Here, we will discuss the case of 
the kaon as an
example. The case of the pion is completely analogous.

\subsection{Two-point function analysis}\label{sec:2pt}

\begin{figure}
	\includegraphics[width=0.45\textwidth]{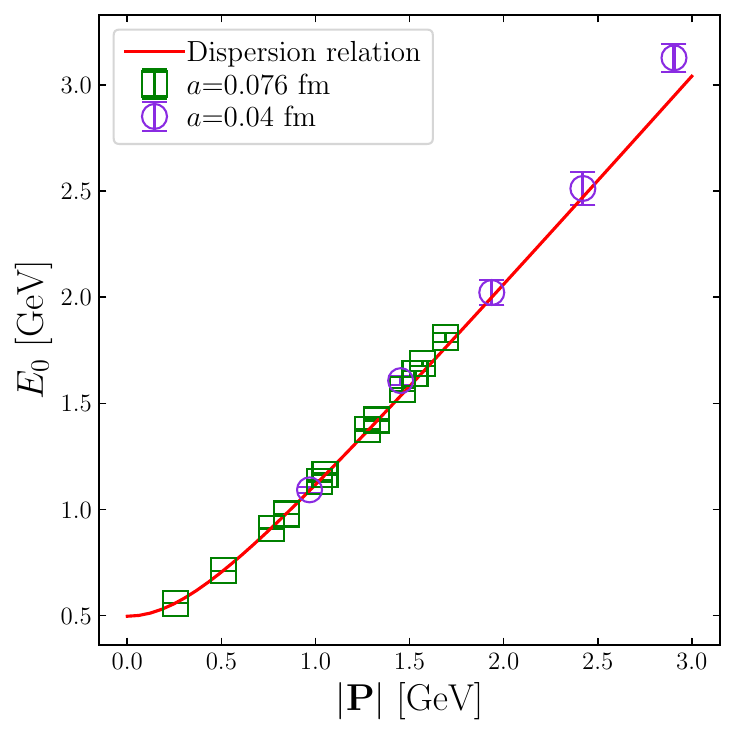}
	\caption{The ground-state energy $E_0$ extracted from the two-point functions on the $a$ = 0.076 fm (squared symbols) and 0.04 fm (circled symbols) lattices are shown. The red line is calculated from the dispersion relation $E=\sqrt{m_K^2+|\textbf{P}|^2}$ with $m_K$ = 0.497 GeV.\label{fig:disp}}
\end{figure}

The two-point functions of kaon can be decomposed as
\begin{align}\label{eq:c2ptsp}
\begin{split}
    C_{\rm 2pt}(\mathbf{P},t_s)&=\sum_{n=0}^{N_{\rm state}-1} |A_n|^2 (e^{-E_n t_s}+e^{-E_n (aL_t-t_s)}).
    \end{split}
\end{align}
Here, $E_{n}$ represents the energy levels, $A_n=\langle\Omega|K_S|n;\mathbf{P}\rangle$ is the 
kaon overlap amplitude, and $|\Omega\rangle$ denotes the vacuum state.
By truncating the spectrum decomposition up to $N$-state, 
we can extract the first few energy levels and overlap amplitudes by fitting the two-point function data. 
Here, we choose $N_{\rm state}=2$ to obtain the energy levels and the overlap amplitudes.
The results on the ground-state energy of the kaon for different momenta are shown in \fig{disp}.
As one can see from the figure, the results on the ground-state energy $E_0$ as a function of $|\textbf{P}|$
agree very well with the line given by the dispersion 
relation $\sqrt{m_K^2+|\textbf{P}|^2}$ with $m_K$ = 0.497 GeV. 
We will use the values of $E_n$ and $A_n$ determined here for the 
following analysis of three-point functions.

\subsection{The ground-state bare matrix elements from three-point function}\label{sec:c3pt}

Similar to the two-point functions, the three-point functions have the spectral decomposition,
\begin{align}\label{eq:c3ptsp}
 \begin{split}
C_{\rm 3pt}(\mathbf{P}^f,\mathbf{P}^i;\tau,t_s)=\sum_{m,n}A_mA_n^*\langle m;\mathbf{P}^f|O_\Gamma|n;\mathbf{P}^i\rangle e^{-(t_s-\tau)E_m^f}e^{-\tau E_{n}^i},
 \end{split}
\end{align}
where the overlap amplitudes $A_n$ as well the energy levels $E_n$ are the same as
for the two-point functions, 
and the ground-state bare matrix element $F^B=\langle 0;\mathbf{P}^f|O_\Gamma|0;\mathbf{P}^i\rangle$ is the bare form factor. We construct a ratio defined as
\begin{align}\label{eq:ratio_supp}
    R^{fi}(\mathbf{P}^f,\mathbf{P}^i;\tau,t_s) \equiv \frac{2\sqrt{E_0^f E_0^i}}{E_0^f+E_0^i} \frac{C_{\rm 3pt}(\mathbf{P}^f, \mathbf{P}^i; \tau,t_s)}{C_{\rm 2pt}(\mathbf{P}^f, t_s)}\times \left[ \frac{C_{\rm 2pt}(\mathbf{P}^i, t_s - \tau)C_{\rm 2pt}(\mathbf{P}^f, \tau)C_{\rm 2pt}(\mathbf{P}^f, t_s)}{C_{\rm 2pt}(\mathbf{P}^f, t_s - \tau)C_{\rm 2pt}(\mathbf{P}^i, \tau)C_{\rm 2pt}(\mathbf{P}^i, t_s)} \right]^{1/2}
\end{align}
to take advantage of the correlation between two-point and three-point functions. In the $t_s\rightarrow\infty$ limit, 
the ratios give the bare matrix elements of the ground state 
$R^{fi}(\mathbf{P}^f,\mathbf{P}^i,\tau\rightarrow \infty,t_s\rightarrow \infty) = F^B_M(Q^2)$. Based on the spectral decomposition formula, several methods can be considered to perform the
$\tau\rightarrow \infty,t_s\rightarrow \infty$ extrapolation of the ratios:
\begin{enumerate}
  \item \textit{Two-state fit.} We truncate the spectral decomposition formula of the two-point functions as well as the three-point functions up to two states ($m,n$ can be 0 and 1), insert them into the ratios, and finally get the formula below. 
  \begin{equation}
	\begin{aligned}
		R^{fi}(\mathbf{P}^f, \mathbf{P}^i; \tau, t_s) 
		&= \left[ \mathcal{O}_{00} + \frac{|A_1^i|}{|A_0^i|}\frac{|A_1^f|}{|A_0^f|}\mathcal{O}_{11} e^{(\Delta E^f - \Delta E^i)\tau} e^{- \Delta E^f t_s} +  \frac{|A_1^i|}{|A_0^i|}\mathcal{O}_{01} e^{- \Delta E^i \tau} + \frac{|A_1^f|}{|A_0^f|}\mathcal{O}_{10} e^{-(t_s-\tau)\Delta E^f} \right] \\
		&\quad\times \sqrt{\frac{1 + \frac{|A_1^i|^2}{|A_0^i|^2}\frac{|A_1^f|^2}{|A_0^f|^2} e^{-(t_s - \tau)\Delta E^i} e^{- \Delta E^f \tau} +  \frac{|A_1^i|^2}{|A_0^i|^2} e^{-(t_s-\tau)\Delta E^i} + \frac{|A_1^f|^2}{|A_0^f|^2} e^{- \Delta E^f \tau}}{1 + \frac{|A_1^i|^2}{|A_0^i|^2}\frac{|A_1^f|^2}{|A_0^f|^2} e^{-(t_s - \tau)\Delta E^f} e^{- \Delta E^i \tau} +  \frac{|A_1^i|^2}{|A_0^i|^2} e^{- \Delta E^i \tau} + \frac{|A_1^f|^2}{|A_0^f|^2} e^{-(t_s-\tau)\Delta E^f} }} \\
		&\quad\times \frac{1}{\sqrt{ 1 + \frac{|A_1^i|^2}{|A_0^i|^2}\frac{|A_1^f|^2}{|A_0^f|^2} e^{-(\Delta E^f + \Delta E^i)t_s} + \frac{|A_1^i|^2}{|A_0^i|^2} e^{- \Delta E^it_s} + \frac{|A_1^f|^2}{|A_0^f|^2} e^{- \Delta E^ft_s} }}.
    \end{aligned}
  \end{equation}
  In the Breit frame limit, we can use a simplified form for this ratio:
  \begin{align}
      R^{fi}(\mathbf{P}^f, \mathbf{P}^i; \tau, t_s) = \frac{\mathcal{O}_{00}+\frac{|A_1|^2}{|A_0|^2}\mathcal{O}_{11}e^{-t_s\Delta E}+\frac{|A_1|}{|A_0|}\mathcal{O}_{01}e^{-\tau\Delta E}+\frac{|A_1|}{|A_0|}\mathcal{O}_{10}e^{-(t_s-\tau)\Delta E}}{1+\frac{|A_1|^2}{|A_0|^2}e^{-t_s\Delta E}},
  \end{align}
with $\Delta E=E_1-E_0$ being the energy difference and 
$\mathcal{O}_{mn} = \langle m;\mathbf{P}^f|O_\Gamma|n;\mathbf{P}^i\rangle$ 
representing the matrix elements, in which $\mathcal{O}_{00}$ is the bare form factor. 
We perform the fit using the above expressions by fixing $\Delta E$ as well as $|A_1|/|A_0|$ to the values
obtained from the analysis of the two-point functions, treating $\mathcal{O}_{nm}$ as fit parameters, 
and skipping $n_{\textup{sk}}$ data points in $\tau$ 
on the two sides of the source-sink separation to avoid excited-state contamination. 
In general, this implies a four-parameter fit. In the case of the Breit frame, we have $\mathcal{O}_{01}=\mathcal{O}_{10}$, so 
we deal with a three-parameter fit. We will refer to this fit method as Fit($n_{\textup{sk}}$). 
Alternatively, instead of fixing $\Delta E$ and $|A_1|/|A_0|$, we could impose priors into the $\chi^2$ by utilizing their central values and corresponding 1-$\sigma$ errors derived from the two-point functions, and perform a constrained five-parameter fit, with $\mathcal{O}_{00},
\mathcal{O}_{01},\mathcal{O}_{11},\Delta E$ and $|A_1|/|A_0|$ being fit parameters. We refer to this fit method as 2-prior Fit($n_{\textup{sk}}$).

  \item \textit{Summation method.} We construct the sum of the ratios over time insertion $\tau$,
  \begin{align}
      R^{fi}_{\textup{sum}}(t_s)=\sum_{\tau=n_{\textup{sk}}a}^{t_s-n_{\textup{sk}}a}R^{fi}(t_s,\tau).
  \end{align}
  For sufficiently large $t_s$, the excited-states contribution is suppressed, and the sum can be approximated by a linear function,
  \begin{align}
  \begin{aligned}
  R^{fi}_{\textup{sum}}(t_s)=nF^B+B_0+\mathcal{O}(e^{-(E_1-E_0)t_s}), \quad n = t_s-(2n_{\textup{sk}} - 1)a.
  \end{aligned}
  \end{align}
  Therefore, we can do a linear fit of data on $R^{fi}_{\textup{sum}}(t_s)$ to extract the bare form factor, $F^B$. 
  One can control the excited-state contamination by choosing different values of $n_{\textup{sk}}$.
  However, for certain cases when $t_s$ is too small, the excited-state contribution cannot be fully ignored. 
  We can include the LO correction term, allowing us to express the ansatz in the Breit frame as
  \begin{align}
  R^{fi}_{\textup{sum}}(t_s)=nF^B+B_0+nB_1e^{-(E_1-E_0)t_s}.
  \end{align}
  These two methods will be denoted as Sum(n$_{\textup{sk}}$) and SumExp($n_{\textup{sk}}$).
  \item When the $t_s$ is large enough, the excited-states contribution is negligible within the statistic error. In this case, the ratios $R^{fi}(t_s,\tau)$ would show $t_s$- and $\tau$-independent plateaus within the errors. One can simply do a plateau fit to get $F^B$. We refer to this method as Plateau($n_{\textup{sk}}$).
\end{enumerate}

\begin{figure}
\centering
	\includegraphics[width=0.45\textwidth]{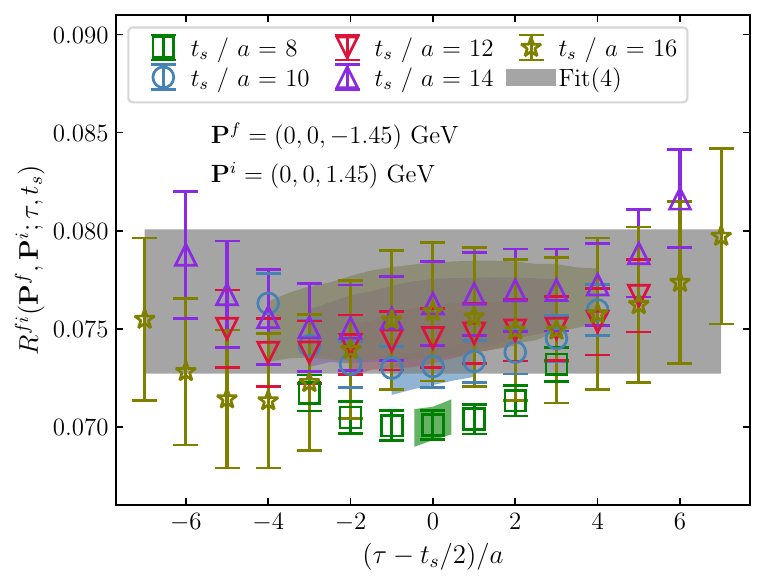}\hspace{0.3cm}
	\includegraphics[width=0.45\textwidth]{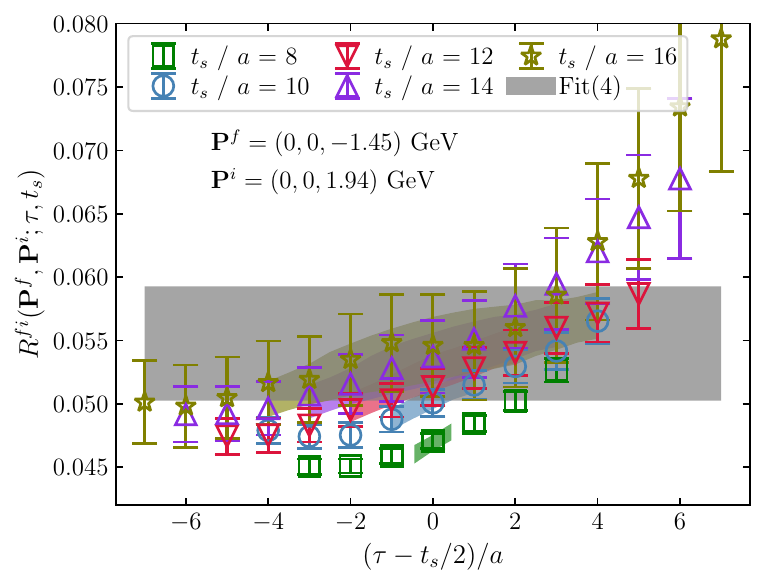}
	\includegraphics[width=0.45\textwidth]{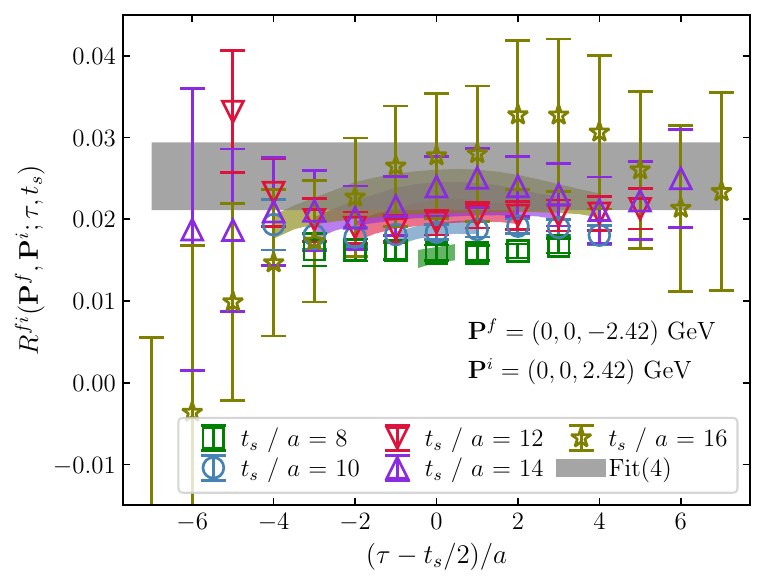}\hspace{0.3cm}
	\includegraphics[width=0.45\textwidth]{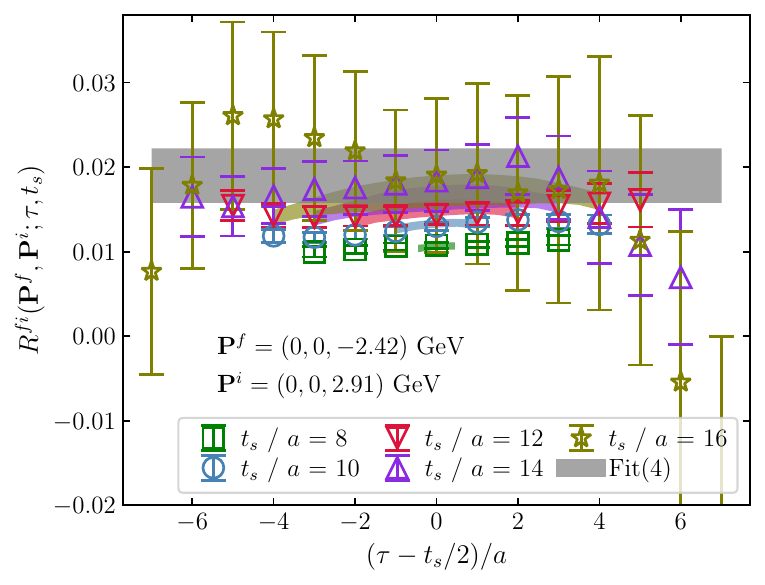}
	\caption{The lattice results of the ratio, $R^{fi}$, and the corresponding two-state Fit($n_{sk}=4$) 
    on $a$ = 0.04 fm lattice are shown. The left panels show examples from the Breit frame, while the right ones are examples from the non-Breit frames. The colored bands present the fit results of the corresponding colored lattice data, while the grey bands show the bare form factor results.
\label{fig:ratio040}} 
\end{figure}

\begin{figure}
\centering
	\includegraphics[width=0.45\textwidth]{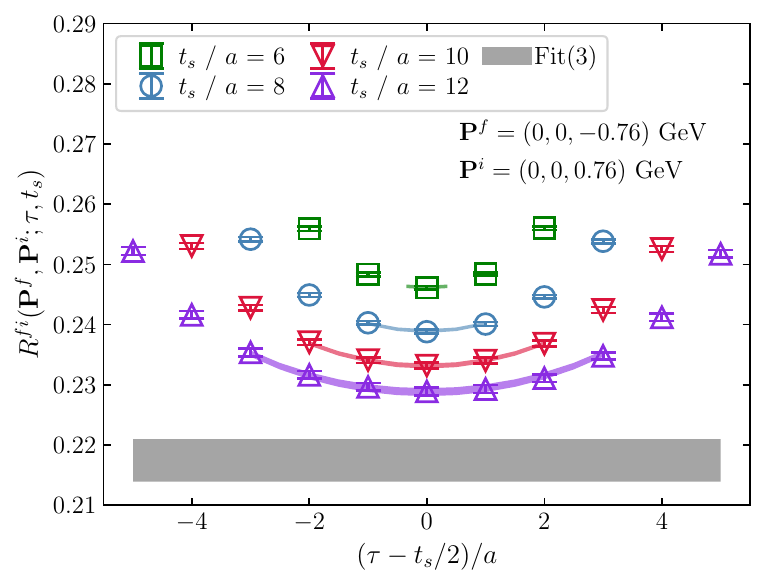}\hspace{0.3cm}
	\includegraphics[width=0.45\textwidth]{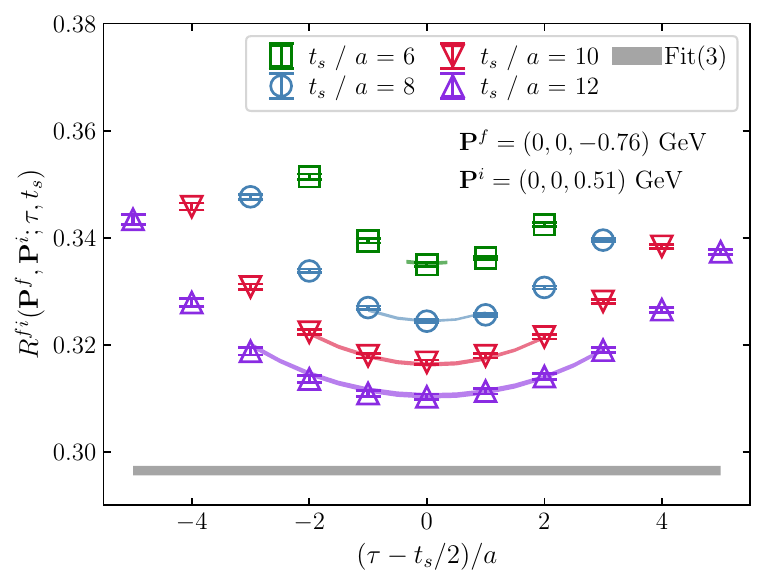}
	\includegraphics[width=0.45\textwidth]{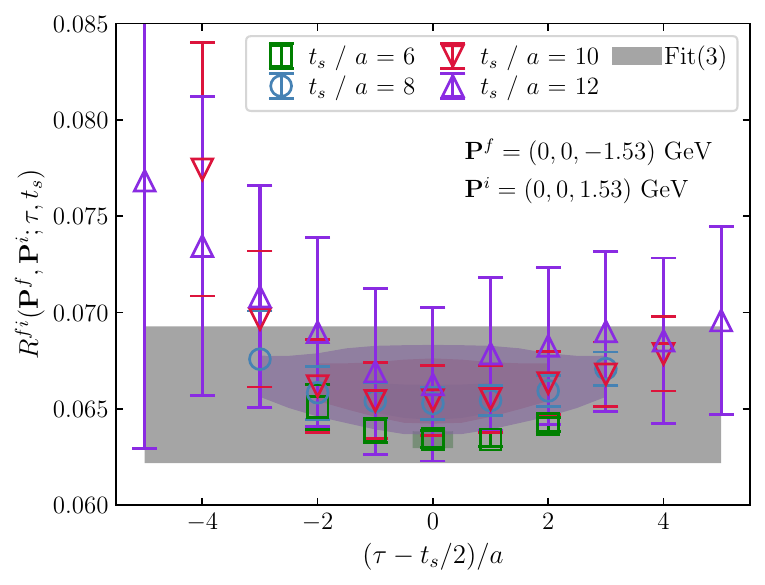}\hspace{0.3cm}
	\includegraphics[width=0.45\textwidth]{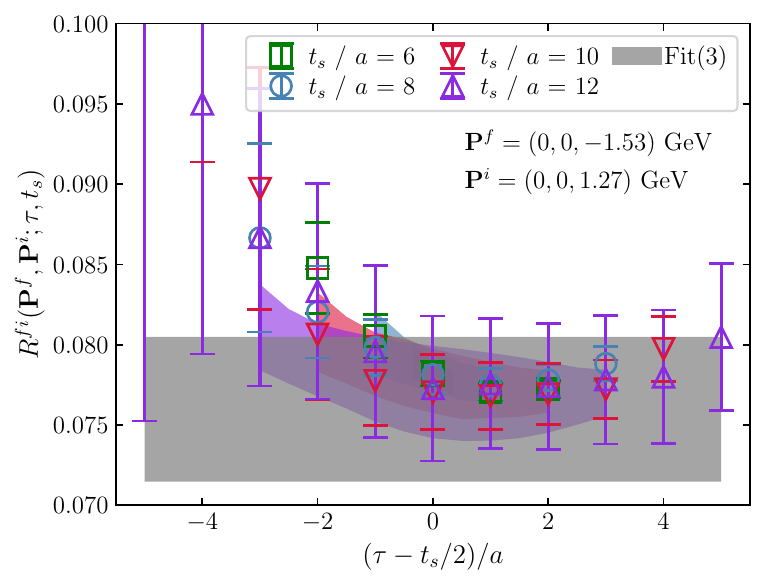}
	\caption{The same as \fig{ratio040}, but for the ratio, $R^{fi}$, on $a$ = 0.076 fm lattice.
\label{fig:ratio076}} 
\end{figure}

In this work, we considered the first two methods. One thing to notice is that $\tau$ has a rather limited range, so the values of $n_{\textup{sk}}$ must be determined appropriately.
Since for $a = 0.04$ fm ensemble, we have source-sink separations 
$t_s/a=\{8,10,12,14,16\}$, the $\tau$-dependence of the three-point correlators can only
be studied for $\tau \lesssim 8a$. Based on our previous experience~\cite{Gao:2021xsm, Gao:2022iex},
here we choose $n_{\textup{sk}}\ge 3$. For $a=0.076$ fm ensemble,  we have $t_s/a=\{6,8,10,12\}$, so the $\tau$-dependence of the three-point function can only be studied for $\tau \lesssim 6 a$. Again, based on past experience, we use $n_{\textup{sk}}\ge 2$ in this case. The two-state fits to our lattice
data are shown in \fig{ratio040}
and \fig{ratio076}. In \fig{ratio040}, we present the examples of the ratio ($R^{fi}$) for the $a$ = 0.04 fm lattice, with $P_z^f=-1.45$ GeV varied from $P_z^i=\{1.45,~1.94\}$ GeV in the upper panels, and $P_z^f=-2.42$ GeV varied from $P_z^i=\{2.42,~2.91\}$ GeV in the lower panels. While \fig{ratio076} show the cases for the $a$ = 0.076 fm lattice, with $P_z^f=-0.76$ GeV, varying $P_z^i=\{0.51,~0.76\}$ GeV in the upper panels, and $P_z^f=-1.53$ GeV, varying $P_z^i=\{1.27,~1.53\}$ GeV in the lower panels. In each panel, the colored bands represent the fit results using the two-state Fit($n_{\rm sk}$) method, while the grey bands denote the extrapolated bare form factor results, $F^B$. The color coding of the bands corresponds to the color
coding of the lattice data for different source-sink separations.

\begin{figure}
\centering
    \includegraphics[width=0.24\textwidth]{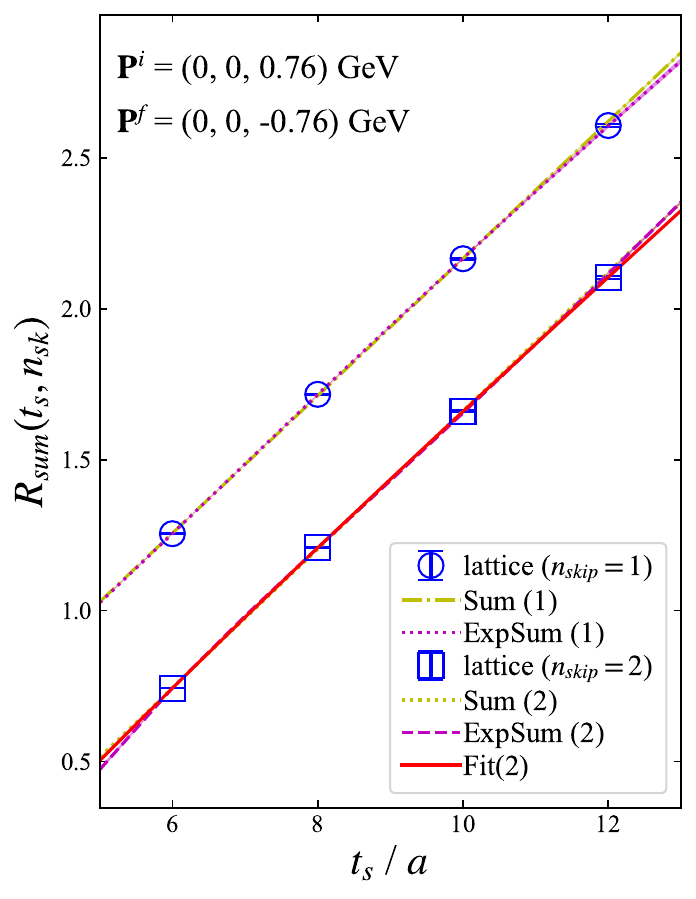}
	\includegraphics[width=0.24\textwidth]{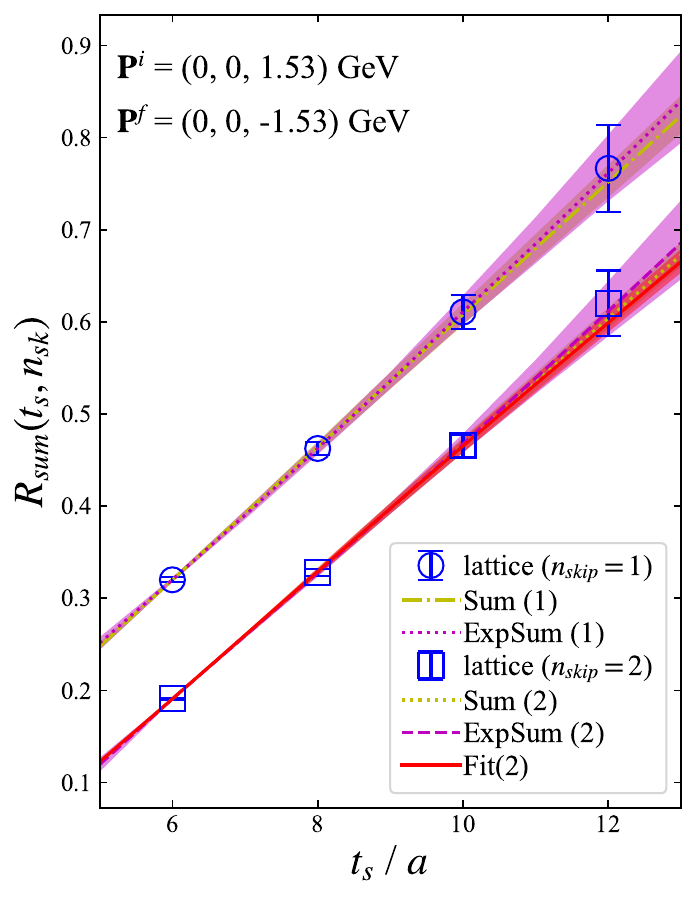}
	\includegraphics[width=0.24\textwidth]{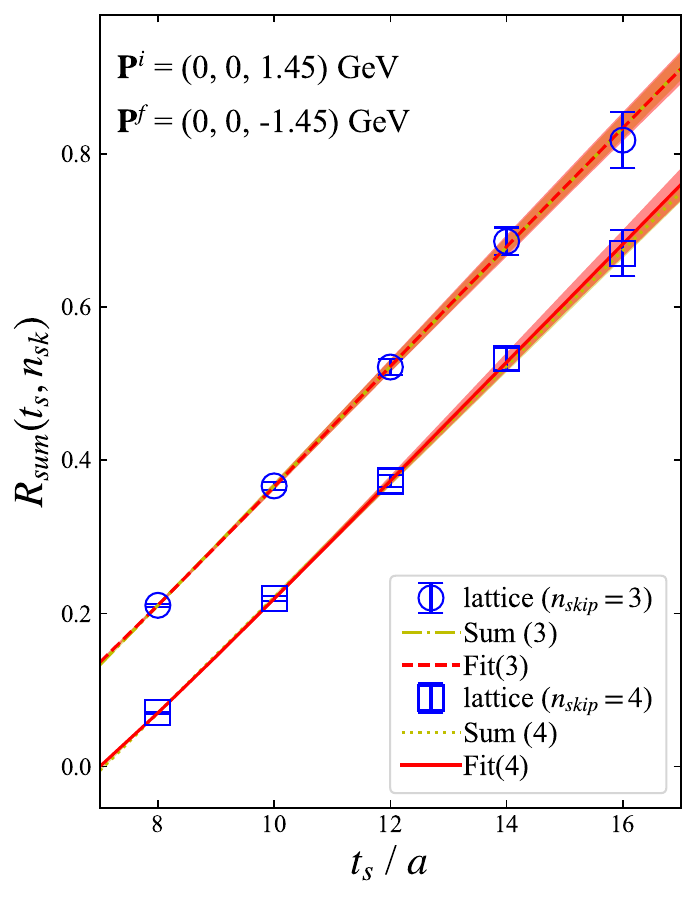}
	\includegraphics[width=0.24\textwidth]{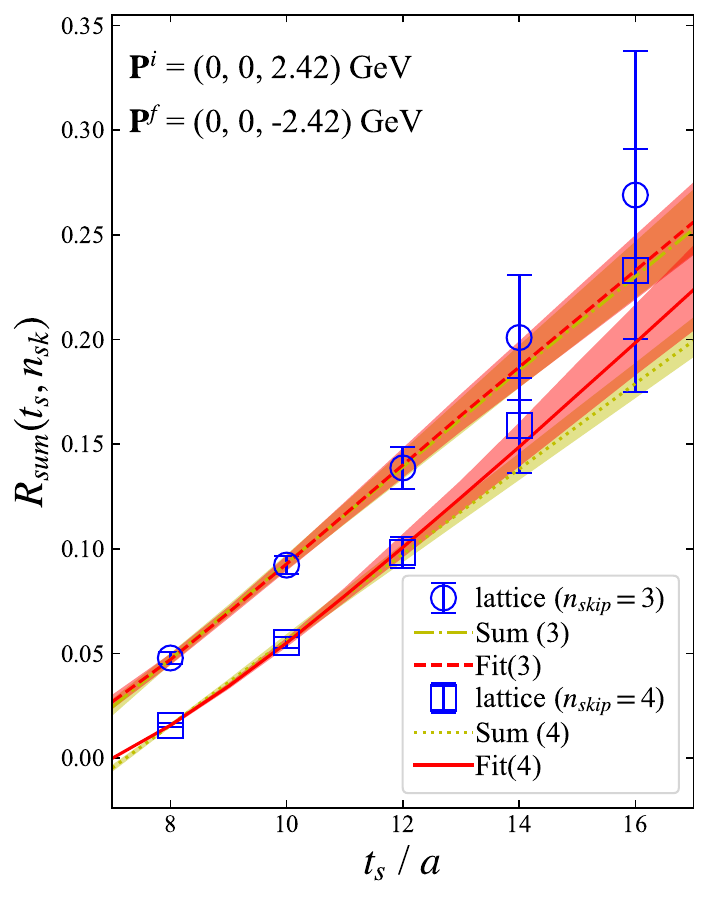}
	\caption{The summation of ratios $R^{fi}_{\textup{sum}}(t_s)$ for $a$ = 0.076 fm (left two panels) and $a$ = 0.04 fm lattice (right two panels) are shown. The curves are reconstructed from the summation fit as well as the two-state fit results.\label{fig:sumfit}}
\end{figure}

\begin{figure}
\centering
    \includegraphics[width=0.46\textwidth]{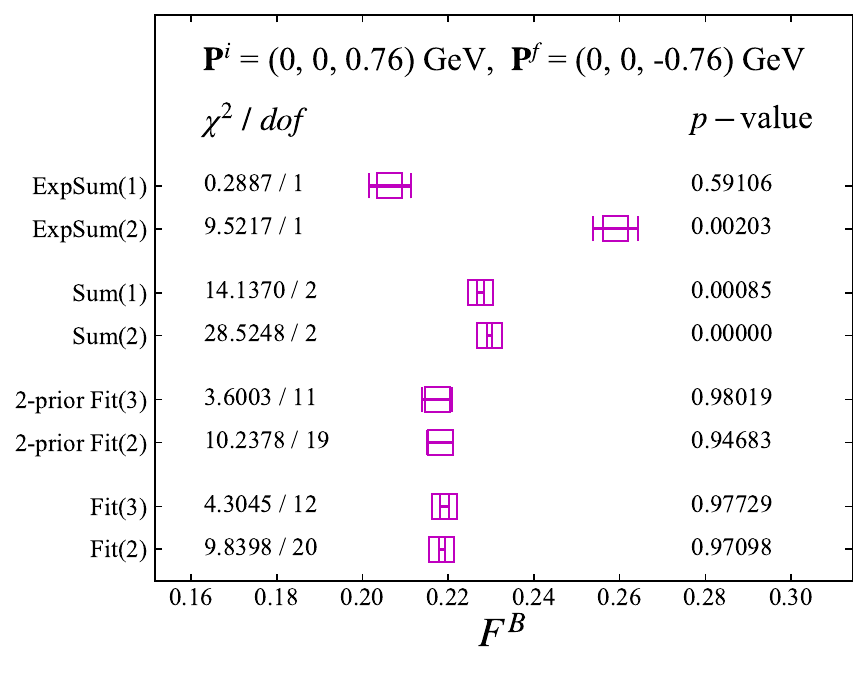}
	\includegraphics[width=0.46\textwidth]{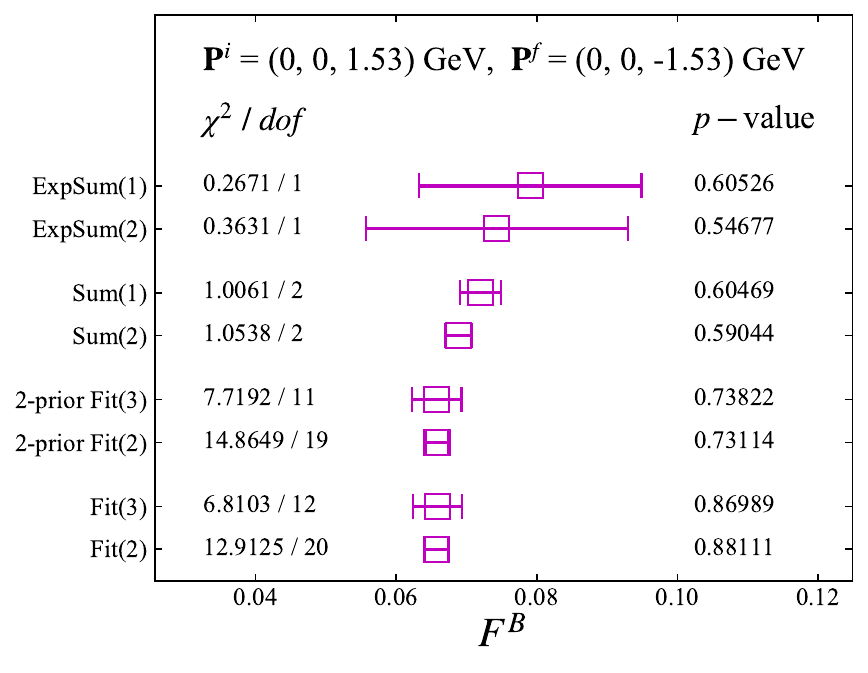} \\
	\hspace{0.35cm}
    \includegraphics[width=0.44\textwidth]{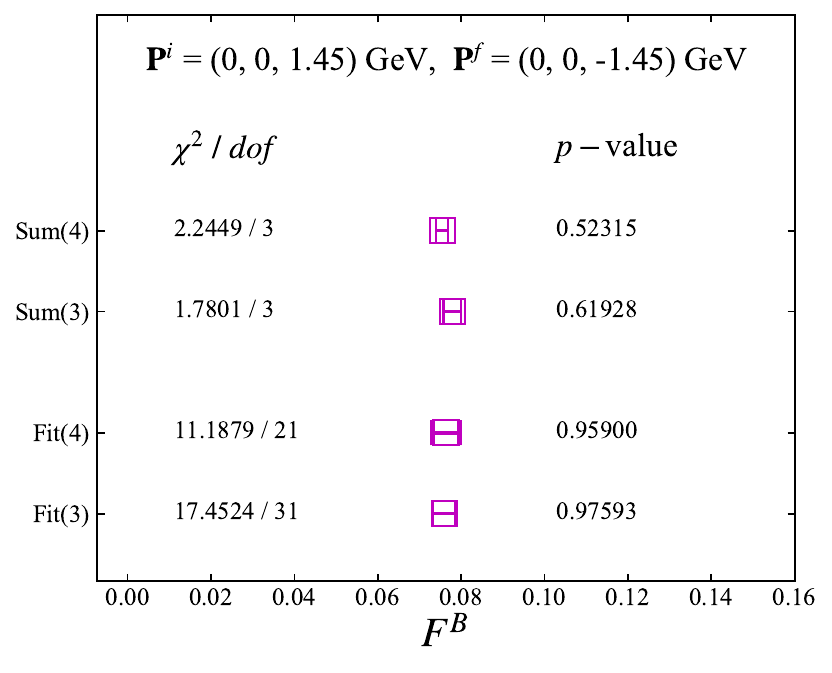}
    \hspace{0.25cm}
	\includegraphics[width=0.43\textwidth]{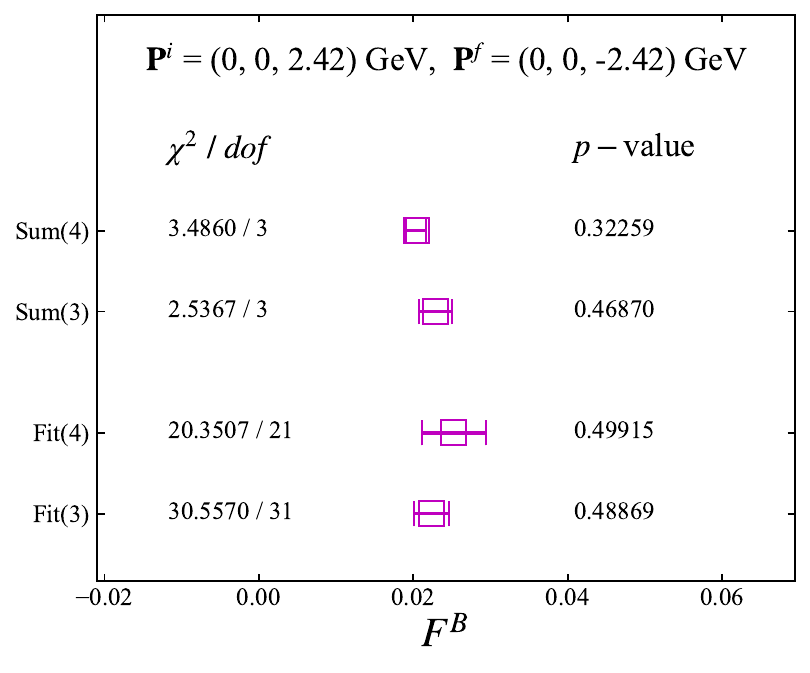}
	\caption{The results and fit quality from the summation fit and two-state fit with different numbers of skipped time insertion $n_{sk}$ are shown for comparison. Upper: $a$ = 0.076 fm lattice, lower: $a$ = 0.04 fm lattice. \label{fig:fitchisq}}
\end{figure}

In \fig{sumfit}, we show the lattice results for $R^{fi}_{\textup{sum}}(t_s)$ 
together with the correspondings fit results from Sum($n_{\textup{sk}}$), ExpSum($n_{\textup{sk}}$), and Fit($n_{\textup{sk}}$) for both $a=0.04$ fm lattice (left two panels) and $a=0.076$ fm lattice (right two panels). 
Furthermore, the values of the bare form factor obtained using these fit methods are summarized in \fig{fitchisq}. 
In this figure, we also show the corresponding 
$\chi^2/dof$ and $p$-value for each method.
As one can see, the two-state fits always give $\chi^2/dof\lesssim$ 1 
and consistent results for different choices of $n_{\textup{sk}}$, 
suggesting this method is robust for the data under consideration. 
For the $a$ = 0.076 fm ensemble and small momenta cases, the summation fits, Sum($n_{\textup{sk}}$), have large values of $\chi^2/dof$.
This is because the statistical errors in these cases are small, and excited-state contamination cannot
be neglected. In this situation, it is 
necessary to introduce the excited-state corrections to the summation method, namely ExpSum($n_{\textup{sk}}$). 
The results from ExpSum($n_{\textup{sk}}$) show reasonable $\chi^2/dof$, but the errors become large since it is
difficult to fit three parameters using only a few data points. On the other hand,
the results from ExpSum($n_{\textup{sk}}$) are consistent with the two-state fit within errors.
For the $a$ = 0.04 fm lattice with larger momenta cases, the excited-state contamination is small 
compared to the statistical errors, so that the results from Sum($n_{\textup{sk}}$) method agree 
with the corresponding results from the two-state fit. Moreover, a comparison of the Breit frame results between the normal Fit and prior Fit methods in the upper panels of \fig{fitchisq} reveals that they are in good agreement with the 1-$\sigma$ error. However, the error magnitude of the prior method can be affected by many factors, like the range of the priors and the statistics, which will sometimes lead to an unreliable error range. Therefore, for the $a=0.076$ fm lattice, we choose the 2-prior Fit(3) results for the Breit frame case and the general Fit(3) results for the non-Breit frame case. For the $a=0.04$ fm lattice, we select the results from general Fit(4) for all cases.

\end{widetext}

\end{document}